\documentclass[aps,prd,preprint,superscriptaddress]{revtex4}
\usepackage{graphicx}
\usepackage{amssymb}

\begin{document}
\newcommand{\vecvar}[1]{\mbox{\boldmath$#1$}}


\begin{flushright}
OU-HET-593, \, December, 2007 \ \ \\
\end{flushright}
\vspace{0mm}

\title{Phenomenological analysis of the nucleon spin contents \\
and their scale dependence}


\author{M.~Wakamatsu and Y.~Nakakoji}
\affiliation{Department of Physics, Faculty of Science, \\
Osaka University, \\
Toyonaka, Osaka 560-0043, JAPAN}



\begin{abstract}
In the past few years, a lot of evidences have been accumulated,
which indicate that the gluon polarization inside the nucleon is likely
to be small at least at the low renormalization scales.
On the other hand, the recent lattice QCD analyses suggest that the net
orbital angular momentum carried by the quarks is nearly zero.
There is also some indication noticed by Brodsky and Gardner based on
the COMPASS observation of small single-spin asymmetry on the isoscalar
deuteron target, that the gluon orbital angular momentum inside the
nucleon is likely to be small. Naively combining all these observations,
we are led to a rather embarrassing conclusion that the nucleon
constituents altogether do not carry enough amount of angular momentum
saturating the total nucleon spin. We show that this somewhat confused
state of affairs can be cleared up only by paying careful attention to
the scale dependencies of the nucleon spin decomposition.

\end{abstract}


\maketitle


\section{Introduction}

If the intrinsic quark spin carries a little of the total nucleon
spin, what carry the rest of it ? This is the famous ``nucleon
spin problem'' raised by the EMC measurements nearly twenty years
ago \cite{EMC88},\cite{EMC89}.
In the past few years, there have been several remarkable progresses
toward the resolution of this long-standing problem.
Firstly, a lot of experimental evidences have been accumulated,
which indicate that the gluon polarization inside nucleon is likely
to be small at least at the low renormalization
scales \cite{COMPASS06}\nocite{PHENIX06}\nocite{STAR05}-\cite{STAR06}.
At the least, it is now widely accepted that the $U_A (1)$-anomaly
motivated explanation of the nucleon spin puzzle is disfavored.
Secondly, the quark spin fraction or the net longitudinal quark polarization
$\Delta \Sigma$ has been fairly precisely determined through the
high-statistics measurements of deuteron spin structure function by the
COMPASS \cite{COMPASSD05},\cite{COMPASSD06} and HERMES
groups \cite{HERMESD06}. According to their new analyses, the portion
of the nucleon spin coming from the intrinsic quark spin is around $30 \%$.
Putting together these two observations blindly, one might be led to the
conclusion that the rest of the nucleon spin must
be carried by the orbital angular momentum of quarks and/or gluons.
On the other hand, however, the recent lattice QCD simulations indicate
that the net orbital angular momentum carried by
the quark fields is very small or close to 
zero \cite{QCDSF04a}\nocite{QCDSF04b}\nocite{LHPC-SESAM03}
\nocite{LHPC-SESAM04}\nocite{LHPC04}\nocite{LHPC05}
\nocite{LHPC07}-\cite{QCDSF-UKQCD07}.
Besides, based on the conjecture on the relation between the
Sivers mechanism and the quark and gluon orbital angular
momenta \cite{Sivers90},\cite{Burk02},
Brodsky and Gardner \cite{BG06} argued that the small single-spin asymmetry
observed by COMPASS collaboration on the
deuteron target \cite{COMPASS-SI05} is an indication of small gluon
orbital angular momentum inside the nucleon.

Naively combining all the observations above, we might be led to the
conclusion that the nucleon constituents on aggregate do not
carry enough amount of angular momentum saturating the total nucleon spin.
What's wrong with the above deduction? The purpose of the present study
is to resolve the apparent paradox above.
To clear up this confused status of our understanding of the
nucleon spin puzzle, we propose to carry out an analysis, in which a
special care is paid to the fact that the decomposition of the nucleon
spin is an absolutely scale-dependent idea. 
What plays a central role in this analysis is Ji's angular momentum
sum rule, supplemented with some additional knowledge listed below.
The first is the information obtained from the recent theoretical
studies of the isoscalar and isovector combinations of the nucleon
anomalous gravitomagnetic moments, $B_{20}^{u+d}(0)$ and 
$B_{20}^{u-d}(0)$, within the lattice QCD as well as within the
chiral quark soliton model (CQSM). The 2nd is the empirical
information on the momentum fractions carried by the quarks and gluons,
as well as on the longitudinal quark polarizations.
The 3rd is the observation, first made by Ji, that the total angular
momentum fractions carried by the quarks and gluons obey exactly the
same evolution equations as the momentum fractions of the quarks
and gluons do.

The plan of the paper is as follows. First, in sect.II, we briefly
review main predictions of the lattice QCD simulations for generalized
form factors and the spin contents of the nucleon carried out
in the past few years. On the other hand, sect III is devoted to new
and improved investigation of the corresponding generalized form
factors within the framework of the chiral quark soliton
model (CQSM). Next, in sect IV, armed with the knowledge gained in
the previous two sections, we try to carry out
semi-empirical analysis of the nucleon spin contents by paying
special attention to their scale dependence. Several concluding remarks
will then be given in sect.V.

\section{Lattice QCD predictions on nucleon spin contents}

Most theoretical analyses of the nucleon spin contents nowadays
heavily relies upon Ji's angular momentum sum
rule \cite{Ji97}\nocite{HJL99}\nocite{JTH96}-\cite{Ji98}.
According to it, the total angular momentum carried by the quark
field with flavor $q$ is given as
\begin{equation}
 J^q \ = \ \frac{1}{2} \,\left[\, A_{20}^q (0) + B_{20}^q (0) \,\right]
 \ = \ \frac{1}{2} \,\left[\, \langle x \rangle^q + B_{20}^q (0) \,\right].
\end{equation}
Here, $A_{20}^q (0)$ is the forward ($t \rightarrow 0$) limit of the
generalized Dirac form factor $A_{20}^q (t)$,
which is related to the 2nd moment of
the unpolarized spin-non-flip generalized parton distribution function
(GPD) $H^q (x,\xi,t)$. It just reduces to the momentum fraction
$\langle x \rangle^q$ carried by the quark with flavor $q$.
On the other hand, $B_{20}^q (0)$ is the forward limit of the
generalized Pauli form factor $B_{20}^q (t)$, which is sometimes
called the anomalous gravitomagnetic moment (AGM).
(More precisely, $B_{20}^q (0)$ is
the contribution of the quark with flavor
$q$ to the nucleon AGM.) The quantity $B_{20}^q (0)$ is also
related to the 2nd moment of the unpolarized spin-flip generalized
parton distribution $E^q (x,\xi, t)$, so that it is in principle
measurable through the high energy deeply virtual Compton
scatterings (DVCS) and/or deeply virtual meson production (DVMP) 
processes \cite{Ji97},\cite{Ji98}.
Confining to the two flavor case, for simplicity, we have two independent
relations : 
\begin{eqnarray}
 J^{u+d} &=& \frac{1}{2} \,\left[\, \langle x \rangle^{u+d} + 
 B_{20}^{u+d} (0) \,\right], \\
 J^{u-d} &=& \frac{1}{2} \,\left[\, \langle x \rangle^{u-d} + 
 B_{20}^{u-d} (0) \,\right].
\end{eqnarray}
Since the quark momentum fraction $\langle x \rangle^{u+d}$ and
$\langle x \rangle^{u-d}$ are empirically known fairly well,
the knowledge of $B_{20}^{u+d} (0)$ and $B_{20}^{u-d} (0)$ is essential
to extract the total angular momentum $J^u$ and $J^d$ carried by the
$u$- and $d$-quarks. In fact, these are the quantities of central
interest in several lattice QCD 
studies \cite{QCDSF04a}\nocite{QCDSF04b}\nocite{LHPC-SESAM03}
\nocite{LHPC-SESAM04}\nocite{LHPC04}\nocite{LHPC05}\nocite{LHPC-SESAM02}
\nocite{LHPC07}-\cite{QCDSF-UKQCD07}.
Here, we briefly review the
relevant predictions of lattice QCD studies on the nucleon spin contents
in the past few years.

We first look into the results on $B_{20}^{u+d} (0)$ and
$B_{20}^{u-d} (0)$ reported by the QCDSF Collaboration in
\cite{QCDSF04a},\cite{QCDSF04b} some years ago. Their predictions are
\begin{equation}
 B_{20}^{u+d} (0) \ = \ 0.102 \pm 0.113, \ \ \ \ \ 
 B_{20}^{u-d} (0) \ = \ 0.566 \pm 0.113. \label{AGMisiv}
\end{equation}
(We recall that their simulations were performed in the so-called
heavy-pion region with $m_\pi \simeq (640 - 1070) \,\mbox{MeV}$.
The values quoted in (\ref{AGMisiv}) are those extrapolated to the
physical pion mass. In practice, however, no strong pion mass
dependencies were observed in their simulations at this stage.)
Combining (\ref{AGMisiv}) with their predictions on $A_{20}^{u+d}(0)$
and $A_{20}^{u-d}(0)$, given by
\begin{equation}
 A_{20}^{u+d} (0) \ = \ \langle x \rangle^{u+d} \ = \ 
 0.547 \pm 0.022, \ \ \ \ \ 
 A_{20}^{u-d} (0) \ = \ \langle x \rangle^{u-d} \ = \
 0.253 \pm 0.022.
\end{equation}
they estimated that
\begin{eqnarray}
 2 \,J^u &=& \,\,0.74 \pm 0.12, \ \ \ \ 
 2 \,J^d \ = \ - \,0.08 \pm 0.08.
\end{eqnarray}
Further combining with their results on the quark polarization,
\begin{eqnarray}
 \Delta u + \Delta d &=& 0.60 \pm 0.02, \ \ \ \ \ 
 \Delta u - \Delta d \ = \ 1.08 \pm 0.02,
\end{eqnarray}
they concluded that the net orbital angular momentum (OAM) of the quarks
is very small or consistent with zero : 
\begin{equation}
 2 \,L^{u+d} \ = \ 0.06 \pm 0.14.
\end{equation}
An independent studies of $B_{20}^{u+d}(0)$ and $B_{20}^{u-d}(0)$
is reported by the LHPC Collaboration 
\cite{LHPC-SESAM03}\nocite{LHPC-SESAM04}\nocite{LHPC04}-\cite{LHPC05} :
\begin{eqnarray}
 B_{20}^{u+d}(0) &=& - 0.09 \pm 0.03, \ \ \ \ \ 
 B_{20}^{u-d}(0) \ = \ \,\,0.67 \pm 0.03.
\end{eqnarray}
Using their previous results for the quark momentum fractions as well as
the quark longitudinal polarizations \cite{LHPC-SESAM02},
\begin{equation}
 \Delta \Sigma \ = \ 0.682 \pm 0.018,
\end{equation}
they also estimated the quark orbital
angular momentum to get
\begin{eqnarray}
 L^u &=& - \,0.088 \pm 0.019, \ \ \ \ \ 
 L^d \ = \ \,\,0.036 \pm 0.013,
\end{eqnarray}
or
\begin{eqnarray}
 2 \,L^{u+d} &=& - \,0.104 \pm 0.038, \ \ \ \ \ 
 2 \,L^{u-d} \ = \ - \,0.248 \pm 0.038,
\end{eqnarray}
Their conclusion at this stage was as follows. Both flavor separately
give a rather small contribution of the order of $17 \,\%$ ($7 \,\%$)
for $u$-quark ($d$-quark) to the nucleon spin, due to cancellation
in quark momentum fraction, spin and $B_{20}$ \cite{LHPC05}.
Adding further $u$ and $d$ contributions give a very small and negative
total orbital angular momentum.

Comparing the results of the two groups, one notices several
discrepancies. For instance, the central value of the QCDSF prediction
for $B_{20}^{u+d}(0)$ is small and positive, while the corresponding
prediction by the LHPC group is small and negative.
In spite of these discrepancies, a main conclusion of the two
analyses was common : the net OAM carried by the quarks is very small
or consistent with zero.
As admitted by themselves, however, a main problem of their analyses
was that these conclusions were obtained from the simulations
performed with fairly large pion mass, ranging from
$640 \,\mbox{MeV}$ to $1070 \,\mbox{MeV}$.

Very recently, both groups carried out more refined analyses
of the nucleon spin contents. The simulations were extended to much
lower pion mass and the results were further extrapolated to the
physical pion mass with the help of chiral perturbation
theory. We first overview the main results of the LHPC
Collaboration \cite{LHPC07}.
For the chiral extrapolation, they tried several versions of chiral
perturbation theory, i.e. covariant baryon chiral perturbation theory
(BChPT), heavy baryon chiral perturbation theory (HBChPT)
with and without the $\Delta$ resonance.
The results obtained with use of covariant BChPT are
\begin{eqnarray}
 A_{20}^{u+d}(0) &=& \,\,0.520 \pm 0.014, \ \ \ \ \ \ 
 A_{20}^{u-d}(0) \ = \ \,\,0.157 \pm 0.006, \\
 B_{20}^{u+d}(0) &=& - \,0.094 \pm 0.050, \ \ \ \ \ 
 B_{20}^{u-d}(0) \ = \ \,\,0.274 \pm 0.037,
\end{eqnarray}
which give
\begin{equation}
 2 \,J^{u+d} \ = \ 0.426 \pm 0.052, \ \ \ 
 2 \,J^u \ = \ 0.428 \pm 0.032, \ \ \ 
 2 \,J^d \ = \ - \,0.002 \pm 0.032 .
\end{equation}
On the other hand, the predictions obtained with the HBChPT without
the $\Delta$ resonance are given only for the isoscalar quantities : 
\begin{eqnarray}
 A_{20}^{u+d}(0) &=& \,0.485 \pm 0.014, \\
 B_{20}^{u+d}(0) &=& \,0.050 \pm 0.049,
\end{eqnarray}
which give
\begin{equation}
 2 \,J^{u+d} \ = \ 0.526 \pm 0.048 .
\end{equation}
One sees that the final answers are fairly sensitive to the ways
of chiral extrapolation. In particular, $B_{20}^{u+d}(0)$, one of our
central interest, is slightly negative in the covariant BChPT,
while it is slightly positive in HBChPT. In either case, combined
with their new preliminary estimate for the quark spin
$\tilde{A}_{10}^{u+d}(t=0) = \Delta \Sigma^{u+d}$, they reconfirmed
their previous conclusion that the net quark orbital angular
momentum is nearly zero.

The QCDSF-UKQCD Collaboration also carried out a similar
analysis \cite{QCDSF-UKQCD07}.
Their main results are summarized as
\begin{eqnarray}
 A_{20}^{u+d}(0) &=& \,\,0.572 \pm 0.012, \ \ \ \ \ \ \,\,\,
 A_{20}^{u-d}(0) \ = \ \,\,0.198 \pm 0.008, \\
 B_{20}^{u+d}(0) &=& - \,0.120 \pm 0.023, \ \ \ \ \ 
 B_{20}^{u-d}(0) \ = \ \,\,0.269 \pm 0.020.
\end{eqnarray}
We point out that these new results by the QCDSF-UKQCD group changed
considerably from the previous QCDSF predictions obtained in the
heavy-pion region several years ago  \cite{QCDSF04a},\cite{QCDSF04b}.
This would mainly be an effect of chiral extrapolation to the physical
pion mass. Putting aside moderate changes of
$A_{20}^{u+d}(0)$ and $A_{20}^{u-d}(0)$, the changes of
$B_{20}^{u+d}(0)$ and $B_{20}^{u-d}(0)$ are drastic. First, even
the sign is changed for $B_{20}^{u+d}(0)$, although the fact, that
its absolute value is relatively small, is intact.
Also drastic is a considerable (more than a factor of two)
reduction of the magnitude of isovector $B_{20}^{u-d}(0)$.
(This is also the case for the old and new LHPC predictions for
$B_{20}^{u-d}(0)$ \cite{LHPC04},\cite{LHPC05},\cite{LHPC07}.)
Here, we emphasize that this reduction was
predicted in our theoretical analysis of $B_{20}^{u-d}(0)$
within the chiral quark soliton model \cite{WakaNaka06}.
In fact, it was shown there that
this quantity has a strong pion mass dependence and that the
lattice QCD predictions
obtained in the heavy-pion region has a danger of overestimating it.
The QCDSF-UKQCD Collaboration also carried out a new estimate of
$\tilde{A}_{10}^{u+d}(0) = \Delta \Sigma^{u+d}$ \cite{QCDSF-UKQCD07}
and obtain
\begin{equation}
 \Delta \Sigma^{u+d} \ = \ 0.402 \pm 0.048.
\end{equation}
Combining these, they finally obtain an estimate
\begin{eqnarray}
 2 \,J^{u+d} &=& 0.452 \pm 0.026, \ \ \ \ \ 
 2 \,L^{u+d} \ = \ 0.050 \pm 0.054.
\end{eqnarray}
Thus, despite some appreciable changes of the predictions for some
generalized form factors, a common conclusion of the two lattice QCD
groups, that the net quark OAM is small, appears to be reconfirmed
also by these new analyses.

Now we have a dilemma. Neither of the intrinsic quark spin, the gluon
polarization, nor the quark OAM seems to carry enough amount of
angular momentum to saturate the total nucleon spin.
Does it mean that the rest of the nucleon spin
is mostly carried by the gluon OAM ? As already mentioned, however,
very large gluon OAM seems to contradict the recent claim
by Brodsky and Gardner based on the observed small single-spin
asymmetry on the deuteron target by the COMPASS
group \cite{BG06},\cite{COMPASS-SI05}.
In our opinion, this confused status arises because we have not
paid enough care to the fact that the decomposition of the nucleon
spin is a highly scale-dependent idea.
Later, we shall carry out an analysis, which pays more
careful attention to the scale dependencies of the nucleon spin
decomposition.

\section{Chiral Quark Soliton Model predictions}

In a previous paper \cite{WakaNaka06}, we investigated the generalized
form factors of the nucleon within the framework of the CQSM.
A particular emphasis was put there on the pion mass dependence of
the relevant quantities. (A similar analysis was carried out also in
\cite{GGOSSU2007A}\nocite{GGOSSU2007B}-\cite{GOSS2006}.
See also \cite{Waka05}, in which the strong pion mass dependence of the
net quark polarization $\Delta \Sigma$ in the chiral region was pointed out.)
We discuss here only the predictions on $A_{20}^{u+d}(0)$,
$B_{20}^{u+d}(0)$, $A_{20}^{u+d}(0)$, and $B_{20}^{u-d}(0)$, which
provide us with enough information for the nucleon spin
decomposition.

A largest discrepancy between the predictions of the CQSM and those
of the lattice QCD simulations was observed for the isovector
AGM $B_{20}^{u-d}(0)$ of the nucleon \cite{WakaNaka06}, so that we
will start our discussion with this quantity.
Within the framework of the CQSM, or more generally in any other
low energy models, the forward ($t \rightarrow 0$) limits
of the isovector Pauli form factor $B_{10}^{u-d} (t)$
as well as the AGM form factor $B_{20}^{u-d} (t)$ are calculated as
the difference of the standard and generalized Sachs magnetic and
electric form factors at $t = 0$ as (see \cite{WakaNaka06} for
more detail)
\begin{eqnarray}
 B_{10}^{u-d} (0) &=& G_{M,10}^{(I=1)} (0) \ - \ 
 G_{E,10}^{(I=1)} (0), \\
 B_{20}^{u-d} (0) &=& G_{M,20}^{(I=1)} (0) \ - \ 
 G_{E,20}^{(I=1)} (0).
\end{eqnarray}
For completeness, we list below the theoretical expressions for
the above quantities within the CQSM.
The isovector electric form factor in the forward limit, i.e.
$G_{E,10}^{(I=1)} (0)$ is just reduced to the isovector charge of
the nucleon, which denotes that
\begin{equation}
 G_{E,10}^{(I=1)} (0) \ = \ 1 .
\end{equation}
On the other hand, the isovector gravitomagnetic moment
$G_{E,20}^{(I=1)}(0)$ is given as
\begin{eqnarray}
 G_{E,20}^{(I=1)}(0) &=& \frac{1}{M_N} \,\frac{1}{3 \,I} \,
 \left( \,\frac{N_c}{2} \,\right) \,
 \sum_{m > 0, n \leq 0} \,\frac{1}{E_m - E_n} \,\,\nonumber \\
 &\,& \times \ 
 \langle m || \,\mbox{\boldmath $\tau$} \,|| n \rangle \,\,
 \left\{ \,\frac{E_m + E_n}{2} \,
 \langle m || \,\mbox{\boldmath $\tau$} \,|| n \rangle \ + \ 
 \langle m || \,\frac{1}{3} \,
 (\mbox{\boldmath $\alpha$} \cdot \mbox{\boldmath $p$}) \,
 \mbox{\boldmath $\tau$} \,|| n \rangle \,\right\} ,
\end{eqnarray}
with $M_N$ being the nucleon mass.
Here, $| n \rangle$ and $E_n$ are the eigenstates and the corresponding
eigenenergies of the static Dirac Hamiltonian $H$ with the hedgehog
mean field, i.e.
\begin{equation}
 H \,| n \rangle \ = \ E_n \,| n \rangle ,
\end{equation}
where
\begin{equation}
 H \ = \ \frac{\mbox{\boldmath $\alpha$} \cdot \nabla}{i} \ + \ 
 \beta \, M \,\left[ \, \cos F(r) \ + \ i \,\gamma_5 \,
 \mbox{\boldmath $\tau$} \cdot \hat{\mbox{\boldmath $r$}} \,
 \sin F(r) \,\right] ,
\end{equation}
with $M$ being the dynamical quark mass. The symbols $\sum_{n \leq 0}$
and $\sum_{m > 0}$ stand for the summation over all the occupied
and unoccupied single-quark eigenstates of $H$. (The fact that
$G_{E,20}^{(I=1)}(0)$ is given as a double sum over the single-quark
orbitals is connected with the fact that it vanishes at the mean-field
level and survives only at the first order in the collective angular
velocity of the soliton.)

Concerning the isovector magnetic moment $G_{M,10}^{(I=1)}(0)$
and the corresponding isovector gravitomagnetic moment
$G_{M,20}^{(I=1)}(0)$, some comments are in order.
In our previous study \cite{WakaNaka06}, we have calculated only
the leading-order contributions to these quantities and neglected the
subleading $1 / N_c$ corrections, for simplicity.
In the present study, we shall include the latters as well.
The reason is because a similar $1 / N_c$ correction (or more concretely,
the 1st-order rotational correction in the collective
angular velocity of the soliton) is known to be important for resolving
the famous underestimation problem of some isovector observables,
like the isovector axial-charge, inherent in the hedgehog-type soliton
model \cite{WW93}\nocite{CBGPPWW94}-\cite{Wakam96}.
Taking account of this 1st order rotational correction,
the isovector magnetic moment of the nucleon consists of the
leading $O (\Omega^0)$ term and the subleading $O (\Omega^1)$ term as
\begin{equation}
 G_{M,10}^{(I=1)}(0) \ = \ G_{M,10}^{(I=1) \,\Omega^0}(0) \ + \ 
 G_{M,10}^{(I=1) \,\Omega^1}(0) ,
\end{equation}
where
\begin{eqnarray}
 G_{M,10}^{(I=1) \,\Omega^0}(0) &=& - \,\frac{M_N}{9} \,N_c \,
 \sum_{n \leq 0} \,\,\langle n || \,
 (\mbox{\boldmath $x$} \times \mbox{\boldmath $\alpha$}) \cdot
 \mbox{\boldmath $\tau$} \,|| n \rangle ,\\
 G_{M,10}^{(I=1) \,\Omega^1}(0) &=& - \,i \,\frac{M_N}{9 \,I} \,
 \left( \,\frac{N_c}{2} \,\right) 
 \sum_{m > 0, n \leq 0} \,\frac{1}{E_m - E_n} \,
 \langle m || \,\mbox{\boldmath $\tau$} \,|| n \rangle \,
 \langle m || \,
 (\mbox{\boldmath $x$} \times \mbox{\boldmath $\alpha$}) \times
 \mbox{\boldmath $\tau$} \,|| n \rangle .
\end{eqnarray}
Similarly, $G_{M,20}^{(I=1)}(0)$ is given as a sum of the
$O (\Omega^0)$ and the $O (\Omega^1)$ terms : 
\begin{equation}
 G_{M,20}^{(I=1)}(0) \ = \ G_{M,20}^{(I=1) \,\Omega^0}(0) \ + \ 
 G_{M,20}^{(I=1) \,\Omega^1}(0) ,
\end{equation}
where
\begin{eqnarray}
 G_{M,20}^{(I=1) \Omega^0}(0) &=& - \,\frac{1}{9} \,N_c \,
 \sum_{n \leq 0} \,
 \left\{\, E_n \,\langle n || \,
 (\mbox{\boldmath $x$} \times \mbox{\boldmath $\alpha$}) \cdot
 \mbox{\boldmath $\tau$} \,|| n \rangle  \ + \ 
 \langle n || \,\mbox{\boldmath $L$} \cdot \mbox{\boldmath $\tau$} \,
 || n \rangle \,\right\},\\
 G_{M,20}^{(I=1) \Omega^1}(0) &=& - \,i \,\,\frac{1}{9 \,I} \,
 \left( \,\frac{N_c}{2} \,\right) \,
 \sum_{m > 0, n \leq 0} \,\frac{1}{E_m - E_n} \nonumber \\
 \times &\,& \!\!\!\!\!\!\!
 \langle m || \,\mbox{\boldmath $\tau$} \,|| n \rangle \,\,
 \left\{\, \frac{E_m + E_n}{2} \,\,
 \langle m || \,(\mbox{\boldmath $x$} \times \mbox{\boldmath $\alpha$}) 
 \times \mbox{\boldmath $\tau$} \,|| n \rangle  \ + \ 
 \langle m || \,\mbox{\boldmath $L$} \times 
 \mbox{\boldmath $\tau$} \,|| n \rangle \,\right\} .
\end{eqnarray}

As usual, the above sums over the eigenstates of $H$ can be evaluated
with use of the discretized momentum basis of Kahana and
Ripka \cite{KR84},\cite{KRS84}.
(Some generalization of the Kahana-Ripka basis is necessary for
the evaluation of the $O (\Omega^1)$ terms including double
sums \cite{WY91}.)
Now, we are ready to show the results of our numerical calculation.
Similarly to the analysis reported in \cite{WakaNaka06}, we see the
effect of varying the pion mass $m_\pi$, by fixing the dynamical
quark mass $M$ to be $400 \,\mbox{MeV}$. 
For that purpose, we prepare self-consistent soliton solutions for
seven values of $m_\pi$, i.e. $m_\pi = 0, 100, 200, 300, 400, 500$,
and $600 \,\mbox{MeV}$, within the double-subtraction Pauli-Villars
regularization scheme \cite{KWW99}. 
Favorable physical predictions
will be obtained by using the value $M = 400 \,\mbox{MeV}$ and
$m_\pi = 100 \,\mbox{MeV}$, since this set gives a self-consistent
solution close to the phenomenologically successful one obtained with
$M = 375 \,\mbox{MeV}$ and $m_\pi = 0 \,\mbox{MeV}$ in the
single-subtraction Pauli-Villars regularization
scheme \cite{WK99}\nocite{Waka03}-\cite{Waka-NuTeV05}.
(For the nucleon mass $M_N$, appearing in the above formulas of the
generalized form factors, the theoretical consistency requires us
to use self-consistent soliton masses. Otherwise, fundamental
conservation laws like the momentum sum rule would be violated.
See \cite{WakaNaka06} for the detail.)

\begin{table}[htbp]
\begin{center}
\renewcommand{\baselinestretch}{1.20}
\caption{The CQSM predictions for the isovector magnetic moment of the
nucleon in dependence of the pion mass. See the text for more
detailed explanation.
\label{Table1:Ivmag}}
\vspace{8mm}
\begin{tabular}{cccc}
\hline
\ \ $m_\pi (\mbox{MeV})$ \ \ & \ \ $G_{M,10}^{(I=1) \,\Omega^0}(0)$ \ \ & 
\ \ $G_{M,10}^{(I=1) \,\Omega^1}(0)$ \ \ & \ \ 
$G_{M,10}^{(I=1) \,\Omega^0 + \Omega^1}(0)$ \ \ \\
\hline\hline
0 & 4.12 & 1.14 & 5.26 \\
\hline
100 & 3.41 & 1.24 & 4.64 \\
\hline
200 & 2.89 & 1.39 & 4.28 \\
\hline
300 & 2.69 & 1.53 & 4.21 \\
\hline
400 & 2.67 & 1.66 & 4.33 \\
\hline
500 & 2.72 & 1.80 & 4.52 \\
\hline
600 & 2.73 & 1.99 & 4.72 \\
\hline
\end{tabular}
\end{center}
\renewcommand{\baselinestretch}{1.20}
\end{table}

Table \ref{Table1:Ivmag} shows the theoretical predictions for the
isovector magnetic moment of the nucleon, in dependence of the
pion mass $m_\pi$. The 2nd and the 3rd columns respectively
stand for the $O (\Omega^0)$ and the $O (\Omega^1)$ contributions,
while their sums are shown in the 4th column.
One can convince that the 1st order rotational correction is very
important for this isovector observables. With the favorable set of
parameters, i.e. $m_\pi = 100 \,\mbox{MeV}$ with $M = 400 \,\mbox{MeV}$,
the theory gives $\mu_p - \mu_n = G_{M,10}^{(I=1)}(0) \simeq 4.64$,
which is remarkably close to the empirically known isovector
magnetic moment of the nucleon :
\begin{equation}
 (\mu_p - \mu_n)^{exp} \ = \ 4.70589 .
\end{equation}

\begin{table}[htbp]
\begin{center}
\renewcommand{\baselinestretch}{1.20}
\caption{The CQSM predictions fo the isovector AGM of the nucleon
in dependence of the pion mass. See the text for more detailed
explanation. \label{Table2:IvAGM}}
\vspace{6mm}
\begin{tabular}{ccccccc}
\hline
$m_\pi (\mbox{MeV})$ \ & \ $G_{M,20}^{(I=1)}(0)$ \ & \  
$G_{E,20}^{(I=1)}(0)$ \ & \ $B_{20}^{(u-d) \,\Omega^0}(0)$ \ & \  
$B_{20}^{(u-d) \,\Omega^1}(0)$ \ & \ $B_{20}^{u-d}(0)$ \ & \ 
$B_{20}^{u-d}(0)$ at $Q^2 = 4 \,\mbox{GeV}^2$ \\
\hline\hline
0 & 0.361 & 0.228 & 0.133 & 0.272 & 0.405 & 0.256 \\
\hline
100 & 0.392 & 0.276 & 0.116 & 0.342 & 0.458 & 0.289 \\
\hline
200 & 0.452 & 0.327 & 0.125 & 0.429 & 0.554 & 0.350 \\
\hline
300 & 0.519 & 0.350 & 0.169 & 0.491 & 0.660 & 0.418 \\
\hline
400 & 0.579 & 0.354 & 0.225 & 0.534 & 0.759 & 0.480 \\
\hline
500 & 0.640 & 0.347 & 0.293 & 0.567 & 0.860 & 0.544 \\
\hline
600 & 0.716 & 0.328 & 0.388 & 0.600 & 0.988 & 0.625 \\
\hline
\end{tabular}
\end{center}
\renewcommand{\baselinestretch}{1.20}
\end{table}

Next, shown in Table \ref{Table2:IvAGM} are the predictions of the CQSM
for relevant generalized form factors in the forward limit, as functions
of $m_\pi$, which are necessary to evaluate the isovector AGM.
The 2nd and the 3rd columns of this table respectively stand for
the isovector gravitomagnetic moment and the gravitoelectric
moment, while the 4th column represents
the leading-order contribution to the isovector AGM of the nucleon.
Note that the numbers in the 4th column are obtained as the difference
of those in the 2nd and the 3rd columns, according to the formula,
$B_{M,20}^{(u-d) \,\Omega^0}(0) \equiv G_{M,20}^{(I=1)}(0) - 
G_{E,20}^{(I=1)}(0)$.
These are the predictions already given in our previous paper.
What is new here is the 5th column, which represent the 1st order
rotational correction to the isovector AGM of the nucleon.
We have already seen that the 1st order rotational correction is
very important for reproducing the observed isovector magnetic moment
of the nucleon. Table \ref{Table2:IvAGM} shows
that the effect of the 1st order rotational correction is even more drastic
for the isovector AGM of the nucleon. This is because the leading-order
estimate of the isovector AGM shown in the 3rd column is obtained as
the difference of the two quantities
$G_{M,20}^{(I=1)}(0)$ and $G_{E,20}^{(I=1)}(0)$, having the same size of
magnitude, and a sizable cancellation occurs between them.
As a consequence, the final predictions of
the CQSM for the isovector AGM of the nucleon, given in the 7th
column are nearly a factor of 3 or 4 larger than our previous
results neglecting the 1st order rotational correction.

At this stage, one might be interested in a comparison with the
predictions of lattice QCD.
One must be careful here. Different from the anomalous magnetic moment
of the nucleon, which is scale independent due to the conservation of
the electromagnetic current, the anomalous gravitomagnetic moment
is a scale-dependent quantity. The predictions of the lattice QCD
simulations corresponds to the renormalization scale of
$Q^2 \simeq 4 \,\mbox{GeV}^2$, while the predictions of the CQSM is
thought to correspond to much lower energy scale around
$Q^2 = 0.30 \,\mbox{GeV}^2$. Fortunately, by making use of Ji's
observation that $J^q$ and $\langle x \rangle^q$ obey exactly the
same evolution equation, we can figure out the scale dependence of
$B_{20}^{u-d}(0)$. (See the next section, for more detail.)
From the predictions of the CQSM for $B_{20}^{u-d}(0)$ given in the
6th column, we have estimated the corresponding values at
$Q^2 = 4 \,\mbox{GeV}^2$. The results are shown in the 7th column
of Table \ref{Table2:IvAGM}. For the favorable pion mass parameter
$m_\pi = 100 \,\mbox{MeV}$, our estimate gives
$B_{20}^{u=d}(0) \simeq 0.289$, which should be compared with the
corresponding prediction $B_{20}^{u-d}(0) = 0.274 \pm 0.037$ of
the new LHPC lattice simulation \cite{LHPC07},
and $B_{20}^{u-d}(0) = 0.269 \pm 0.020$
of the QCDSF-UKQCD one \cite{QCDSF-UKQCD07}.
One finds that the predictions of the CQSM and those of the lattice
QCD simulations are now remarkably close to each other.
This is a welcome result, since it is thought to give a strong
support to the reliability of the theoretical predictions on the
isovector AGM of the nucleon $B_{20}^{u-d}(0)$.

\begin{figure}[htb] \centering
\begin{center}
 \includegraphics[width=14.0cm]{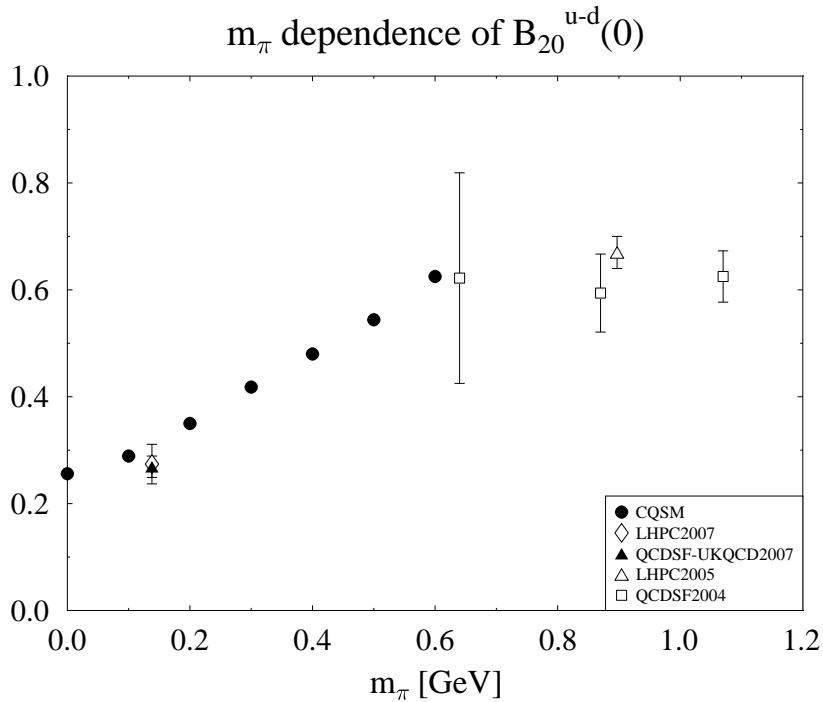}
\end{center}
\vspace*{-0.5cm}
\renewcommand{\baselinestretch}{1.20}
\caption{The pion mass dependence of $B_{20}^{u-d}(0)$
predicted by the CQSM, in comparison with the old and new
lattice QCD predictions. Both correspond to the scale
$Q^2 \simeq 4 \,\mbox{GeV}^2$}
\label{Fig1:IvB20}
\end{figure}%

After obtaining a refined estimate of $B_{20}^{u-d}(0)$ within
the framework of the CQSM, we revise Fig.5(b) in our previous
paper \cite{WakaNaka06}.
The filled circles in Fig.1 are the CQSM predictions
of $B_{20}^{u-d}(0)$ corresponding to the scale
$Q^2 = 4 \,\mbox{GeV}^2$, in dependence of the pion mass.
The corresponding predictions of the QCDSF and LHPC collaborations
carried out in the heavy pion region several years ago
are represented by the open squares and the open
triangles \cite{QCDSF04a},\cite{LHPC05}.
On the other hand, the new predictions by the LHPC and QCDSF-UKQCD
collaborations, extrapolated to the physical pion mass
by utilizing the chiral perturbation theory, are shown respectively
by the open diamond and the filled 
triangles \cite{LHPC07},\cite{QCDSF-UKQCD07}.
One sees that effect of chiral extrapolation is drastic such that
the new predictions of the lattice QCD are more than
a factor of two smaller than the old predictions given in the
heavy pion region. This sizable reduction is just consistent
with our analysis based on the CQSM \cite{WakaNaka06}.
Now, one can convince that the predictions of the CQSM and the lattice
QCD for the isovector AGM of the nucleon $B_{20}^{u-d}(0)$ is mutually
consistent.

Next, we turn to the discussion of more difficult isoscalar quantities.
As shown in \cite{WakaNaka06},\cite{GGOSSU2007A},
the CQSM predicts that
\begin{eqnarray}
 A_{20}^{u+d}(0) &=& \langle x \rangle^{u+d} \ = \ 1, \ \ \ \ \ 
 B_{20}^{u+d}(0) \ = \ 0.
\end{eqnarray}
It should be noticed that these equalities hold irrespectively
of the pion mass within the model.
The 1st relation is only natural. It simply means that the momentum
sum rule is saturated by the quark fields alone in this effective
quark model, which does not contain explicit gluon degrees of
freedom. The 2nd relation holds by the similar reason.
From Ji's angular momentum sum rule, we generally have
(in the two flavor case)
\begin{equation}
 2 \,(J^{u+d} + J^g) \ = \ 
 \langle x \rangle^{u+d} \ + \ B_{20}^{u+d}(0) \ + \ 
 \langle x \rangle^g \ + \ B_{20}^g (0) \ = \ 1.
\end{equation}
If this is combined with the momentum sum rule of QCD,
\begin{equation}
 \langle x \rangle^{u+d} \ + \ \langle x \rangle^g  \ = \ 1,
\end{equation}
we are led to a novel identity,
\begin{equation}
 B_{20}^{u+d}(0) \ + \ B_{20}^{g}(0) \ = \ 0.
\end{equation}
which dictates that the total nucleon AGM (quark plus gluon
contributions) vanishes identically.
The answer $B_{20}^{u+d}(0) = 0$ is therefore an inevitable conclusion
of any effective quark model without gluon fields.
In both of the LHPC and QCDSF lattice QCD simulations carried out
in the heavy-pion region several years ago, the magnitude of
$B_{20}^{u+d}(0)$ was found to be fairly 
small \cite{QCDSF04b},\cite{LHPC-SESAM03}. Since $B_{20}^{u+d}(0)$
is equal to the difference of $2 \,J^{u+d}$ and 
$\langle x \rangle^{u+d}$, the small values of the lattice QCD
predictions for $B_{20}^{u+d}(0)$ at this point were interpreted to
indicate approximate equality of the total angular momentum and
linear momentum fractions of quarks and gluons as advocated by Teryaev
several years ago \cite{Teryaev98}\nocite{Teryaev99}-\cite{Teryaev03}.
However, the recently performed ChPT
fits by the LHPC and QCDSF-UKQCD Collaborations appears to indicate
a sizable bending through the chiral extrapolation in the low
pion mass region, leading to negative $B_{20}^{u+d}(0)$ of the order
of $- \,0.1$, although one must be very careful about the fact that
the final conclusion depends on the ways of chiral extrapolation
method \cite{LHPC07},\cite{QCDSF-UKQCD07}.

Under such circumstances, it would be fine if we can give some
useful constraint on the magnitude of $B_{20}^{u+d}(0)$.
To this end, we first recall the fact that $B_{20}^{u+d}(0)$ is
given as the 2nd moment of the forward limit of the unpolarized
spin-flip GPD $E^{u+d}(x,\xi,t)$ as
\begin{equation}
 B_{20}^{u+d}(0) \ = \ \int_{-1}^1 \,x \,E^{u+d}(x,0,0) \,dx .
\end{equation}
It is important to recognize that the 1st moment of the same quantity
gives the isoscalar magnetic moment of the nucleon up to a factor of 3 :
\begin{equation}
 B_{10}^{u+d}(0) \ = \ \int_{-1}^1 \,E^{u+d}(x,0,0) \,dx \ = \ 
 \kappa^{u+d} \ = \ 3 \,(\kappa^p + \kappa^n).
\end{equation}
The forward limit of the GPD $E^{u+d}(x,0,0)$ was calculated
within the framework of the CQSM by 
Ossmann et al. \cite{Ossmann05}. (There is also a calculation for
the forward limit of the isovector GPD $E^{u-d}(x,0,0)$ within
the CQSM \cite{WT05}.)
It is given as a sum of the two part, i.e. the contribution of
$N_c \,(=3)$ valence quarks and that of the vacuum-polarized
Dirac-sea  quarks as
\begin{equation}
 E^{u+d}(x,0,0) \ = \ E_{val}^{u+d}(x,0,0) \ + \ 
 E_{v.p.}^{u+d}(x,0,0) .
\end{equation}
An interesting findings there are that the valence quark term turns out
to have a similar shape as the corresponding valence term
$f_{val}^{u+d}(x)$ of the standard unpolarized PDF,
while the deformed Dirac-sea contribution has
a strong chiral enhancement near $x=0$, which is antisymmetric
with respect to the transformation $x \rightarrow - \,x$.
(Note that the antisymmetric nature of the Dirac-sea contribution to
$E^{u+d}(x,0,0)$ means that it gives no contribution to its 1st
moment.)
Following the schematic analysis carried out in \cite{GPV01}
(see also \cite{WakaNaka06}),
we therefore propose to parameterize the characteristic feature of
$E^{u+d}(x,0,0)$ in the following simple form : 
\begin{equation}
 E^{u+d}(x,0,0) \ = \ C \,f_{val}^{u+d}(x) \ - \ D \,\delta^\prime (x),
\end{equation}
with $C < 0$, and $D > 0$. With this schematic parameterization,
the 1st and the 2nd moment sum rules of $E^{u+d}(x,0,0)$ become
\begin{eqnarray}
 \int_{-1}^1 \,E^{u+d}(x,0,0) \,dx &=& 3 \,C \ = \ 
 3 \,(\kappa^p \ + \ \kappa^n), \\
 \int_{-1}^1 \,x \,E^{u+d}(x,0,0) \,dx &=& 
 C \,\int_{-1}^1 \,f_{val}^{u+d}(x) \,dx \ + \ D.
\end{eqnarray}
Using the observed anomalous magnetic moments of the proton and the
neutron, the 1st relation gives
\begin{equation}
 C \ = \ (\kappa^p \ + \ \kappa^n)^{exp} \ = \ - \,0.120 .
\end{equation}
On the other hand, the 2nd relation gives
\begin{equation}
 B_{20}^{u+d}(0) \ = \ C \,\int_{-1}^1 \,f_{val}^{u+d}(x) \,dx
 \ + \ D . \label{schematic}
\end{equation}
As a matter of course, in the CQSM, the valence and the vacuum
polarization contributions in (\ref{schematic}) exactly cancel each
other so that the identity $B_{20}^{u+d}(0) = 0$ holds.
Such an exact cancellation may not happen
in real QCD, which contains the gluon fields as well. 
Nonetheless, it is reasonable to expect that the general shape
of $E^{u+d}(x,0,0)$ predicted by the CQSM, especially its chiral
behavior observed in the small $x$ region, would be preserved when
going to real QCD, which in turn strongly indicates that there will be
no change of sign in the contribution of
the sea-quark-like component to $B_{20}^{u+d}(0)$.
We thus conjecture that the coefficient $D$ in (\ref{schematic}) is
at least larger than or equal to $0$.
Combining this with the fact that $\int_{-1}^1 \,f_{val}^{u+d}(x)$ is
smaller than 1 (this is because the sea-quark-like component also
carries some portion of the total momentum fraction of the
nucleon), we would then conclude from (\ref{schematic}) that the lower
limit of $B_{20}^{u+d}(0)$ is $- \,0.12 \ (\,= (\kappa^p + \kappa^n)^{exp})$.
In carrying out a semi-empirical analysis of the nucleon spin contents
in the next section, we therefore take a standpoint that the
precise value of $B_{20}^{u+d}(0)$ is still uncertain, but it
lies most likely in the range
\begin{equation}
 - \,0.12 \ \leq \ B_{20}^{u+d}(0) \ \leq \ 0.
\end{equation}
This is the main theoretical uncertainty in our semi-phenomenological
analysis of the nucleon spin contents carried out in the next section.

\section{Semi-empirical estimate of nucleon spin contents}

Now, we are ready to start our semi-empirical analysis of
the nucleon spin contents. Our strategy here is to use empirical
information as much as possible, if available. 
To explain our approach more concretely,
we start again with Ji's angular momentum sum rule
written in a slightly more general form : 
\begin{equation}
 \frac{1}{2} \ = \ J^Q + J^g,
\end{equation}
with
\begin{eqnarray}
 J^Q &=& \frac{1}{2} \,\left[\,
 \langle x \rangle^Q \ + \ B_{20}^Q(0) \,\right], \ \ \ \ 
 J^g \ = \ \frac{1}{2} \,\left[\,
 \langle x \rangle^g \ + \ B_{20}^g(0) \,\right].
\end{eqnarray}
Here, $Q$ denotes the sum of all active quark flavors.
($Q = u + d$ for the two flavor case, and $Q = u + d + s$ for
the three flavor case.)
To carry out flavor decomposition of the total quark angular
momentum, we also need another combination of Ji's sum rule : 
\begin{equation}
 J^{u-d} \ = \ \frac{1}{2} \,\left[\, \langle x \rangle^{u-d}
 + B_{20}^{u-d} (0) \,\right] .
\end{equation}
We emphasize again that the momentum fractions $\langle x \rangle^Q$,
$\langle x \rangle^{u-d}$, and $\langle x \rangle^g$ are all
empirically well determined.
Naturally, these momentum fractions are
all scale-dependent quantities.
A key observation here, first made by Ji \cite{Ji97},\cite{Ji98},
is that $J^Q$ and $J^g$ obey exactly the same evolution equations as
$\langle x \rangle^Q$ and $\langle x \rangle^g$ do.
According to him, the underlying reason is that forming spatial
moment of energy momentum operator
does not change short distance singularity of the operator.
The solution of this (coupled) evolution equation is extremely
simple at the leading order (LO) : 
\begin{eqnarray}
 2 \,J^Q (Q^2) &=& \frac{3 \,n_f}{16 + 3 \,n_f} 
 \ + \ \left( \frac{\ln Q_0^2 / \Lambda^2}{\ln Q^2 / \Lambda^2} 
 \right)^{2 \,
 (16 + 3 \,n_f) / (33 - 2 \,n_f)} \,
 \left[\,2 \,J^Q (Q_0^2) - \frac{3 \,n_f}{16 + 3 \,n_f} \,\right], \ \ 
 \label{LOJQ} \\
 2 \,J^g (Q^2) &=& \frac{16}{16 + 3 \,n_f}
 \ + \ \left( \frac{\ln Q_0^2 / \Lambda^2}{\ln Q^2 / \Lambda^2}
 \right)^{2 \,
 (16 + 3 \,n_f) / (33 - 2 \,n_f)} \,
 \left[\,2 \,J^g (Q_0^2) - \frac{16}{16 + 3 \,n_f} \,\right] \,. \ \ 
 \label{LOJg}
\end{eqnarray}
Particularly interesting here are the asymptotic values in the
$Q^2 \rightarrow \infty$ limit : 
\begin{equation}
 2 \,J^Q (\infty) \ = \ \frac{3 \,n_f}{16 + 3 \,n_f}, \ \ \ \ 
 2 \,J^g (\infty) \ = \ \frac{16}{16 + 3 \,n_f} .
\end{equation}
Numerically, we obtain
\begin{equation}
 2 \,J^Q (\infty) \ \simeq \ 0.529, \ \ \ \ 
 2 \,J^g (\infty) \ \simeq \ 0.471,
\end{equation}
for $n_f =6$, while
\begin{equation}
 2 \,J^Q (\infty) \ \simeq \ 0.360, \ \ \ \
 2 \,J^g (\infty) \ \simeq \ 0.640, 
\end{equation}
for $n_f = 3$.

In our actual analysis below, we take account of
the scale dependencies of the relevant quantities by using
the known evolution equations at the next-to-leading order (NLO) for
the momentum fractions, making full use of the fact that
$J^q$ and $\langle x \rangle^q$ (and also $J^g$ and
$\langle x \rangle^g$) obey the same evolution equations.
For the sake of completeness, we write down here the relevant NLO
equations, which we use in the following analysis.
The singlet moments $J^Q$ and $J^g$ (and also $\langle x \rangle^Q$
and $\langle x \rangle^g$) evolve as (see, for example, 
\cite{GRV95}\nocite{GRV96}-\cite{WM96})
\begin{eqnarray}
 \left( \begin{array}{c}
 J^Q (Q^2) \\
 J^g (Q^2) \\
 \end{array} \right) &=& \left\{\,
 \left( 
 \frac{\alpha_S(Q^2)}{\alpha_S(Q_0^2)}\right)^{\lambda_-/2 \beta_0} 
 \,\left[\,\mbox{\boldmath $P$}_- - \frac{1}{2 \,\beta_0}\,
 \frac{\alpha_S (Q_0^2) - \alpha_S (Q^2)}{4 \,\pi} \,
 \mbox{\boldmath $P$}_- \,\mbox{\boldmath $R$} \,
 \mbox{\boldmath $P$}_- \right. \right. \nonumber \\
 &-& \left. \left( \frac{\alpha_S(Q_0^2)}{4 \,\pi} - 
 \frac{\alpha_S (Q^2)}{4 \,\pi} \,
 \left( \frac{\alpha_S(Q^2)}{\alpha_S(Q_0^2)}\right)
 ^{(\lambda_+ - \lambda_-)/ 2 \beta_0} \,\right) \,
 \frac{\mbox{\boldmath $P$}_- \,\mbox{\boldmath $R$} \,
 \mbox{\boldmath $P$}_+}{2 \,\beta_0 + \lambda_+ - \lambda_-} \,
 \right] \nonumber \\
 &\,&  \hspace{20mm} + \hspace{10mm} (+ \longleftrightarrow -)
 \hspace{25mm} \Biggr\} \,
 \left( \begin{array}{c}
 J^Q (Q_0^2) \\
 J^g (Q_0^2) \\
 \end{array} \right) . \label{NLOevolsing}
\end{eqnarray}
Here, $\alpha_S (Q^2)$ is the QCD running coupling constant at
the NLO given by
\begin{equation}
 \alpha_S (Q^2) \ = \ \frac{4 \,\pi}{\beta_0 \,\ln(Q^2 / \Lambda^2)}
 \,\left[\, 1 - \frac{\beta_1 \,\ln \ln (Q^2 / \Lambda^2)}{\beta_0^2 \,
 \ln (Q^2 / \Lambda^2)} \,\right] ,
\end{equation}
with the choice $\Lambda = 0.248 \,\mbox{GeV}$, while $\beta_0 = 
11 - \frac{2}{3} \,n_f$ and $\beta_1 = 102 - \frac{38}{3} \,n_f$
with $n_f$ being the active number of quark flavor. The quantities
$\mbox{\boldmath $R$}$ and $\mbox{\boldmath $P$}_{\pm}$ are defined
by
\begin{eqnarray}
 \mbox{\boldmath $R$} &=& 
 \mbox{\boldmath $\gamma$}^{(1)} \ - \  
 \frac{\beta_1}{\beta_0} \,\mbox{\boldmath $\gamma$}^{(0)}, \\
 \mbox{\boldmath $P$}_{\pm} &=& \pm \,
 \frac{\mbox{\boldmath $\gamma$}^{(0)} - \lambda_{\mp}}
 {\lambda_+ - \lambda_-} ,
\end{eqnarray}
where $\mbox{\boldmath $\gamma$}^{(0)}$ and 
$\mbox{\boldmath $\gamma$}^{(1)}$ are the relevant
anomalous dimension matrices at the LO and NLO, respectively,
given by \cite{FRS77}\nocite{LY81}-\cite{ABY97}
\begin{equation}
 \mbox{\boldmath $\gamma$}^{(0)} \ = \ 
 \left( \begin{array}{cc}
 64/9 & - \,4 \,n_f / 3 \\
 - \,64/9 & 4 \,n_f / 3 \\
 \end{array}
 \right)
\end{equation}
and
\begin{equation}
 \mbox{\boldmath $\gamma$}^{(1)} \ = \ \frac{64}{243} \, 
 \left( \begin{array}{cc}
 367 - 39 \,n_f & - \,\frac{1833}{32} \,n_f \\
 - \,(367 - 39 \,n_f) & \frac{1833}{32} \,n_f \\
 \end{array}
 \right)
\end{equation}
while $\lambda_{\pm}$ are the two eigenvalues of the LO
anomalous dimension matrix $\mbox{\boldmath $\gamma$}^{(0)}$.

On the other hand, the nonsinglet (NS) moments evolve as
\cite{GRV95}\nocite{GRV96}-\cite{WM96}
\begin{eqnarray}
 J_{NS} (Q^2) &=& \left[\,1 + \frac{\alpha_S(Q^2) - \alpha_S(Q_0^2)}{4 \pi}
 \,\left(\,\frac{\gamma_{NS}^{(1)}}{2 \,\beta_0} - 
 \frac{\beta_1 \,\gamma_{NS}^{(0)}}{2 \,\beta_0^2} \,\right)\,\right]
 \,\,\left( \frac{\alpha_S (Q^2)}{\alpha_S (Q_0^2)} \,
 \right)^{\gamma_{NS}^{(0)}/ 2 \beta_0} \,J_{NS} (Q_0^2) , \ \ \ \ \ 
 \label{NLOevolNS}
\end{eqnarray}
with
\begin{equation}
 \gamma_{NS}^{(0)} \ = \ \frac{64}{9}, \ \ \ \ 
 \gamma_{NS}^{(1)} \ = \ \frac{64}{243} \,
 (367 - 39 \,n_f).
\end{equation}
Here, it is understood that, for $n_f = 3$, $J_{NS}$ stands for
either of $J^{(3)} \equiv J^u - J^d$ or $J^{(8)} \equiv 
J^u + J^d - 2 \,J^s$.

Now, we are left with two quantities, $B_{20}^{u+d}(0)$ and
$B_{20}^{u-d}(0)$, which are empirically unknown yet.
Here, one might be tempted to use lattice QCD predictions for those.
In our opinion, however, blind acceptance of the lattice QCD predictions
at the present stage is a little dangerous, especially because there
seems to be large uncertainties in the process of chiral extrapolation.
We proceed slightly more cautiously by taking account also the
information from a phenomenologically successful low energy model
of the nucleon, i.e. the CQSM.

\begin{figure}[htb] \centering
\begin{center}
 \includegraphics[width=14.0cm]{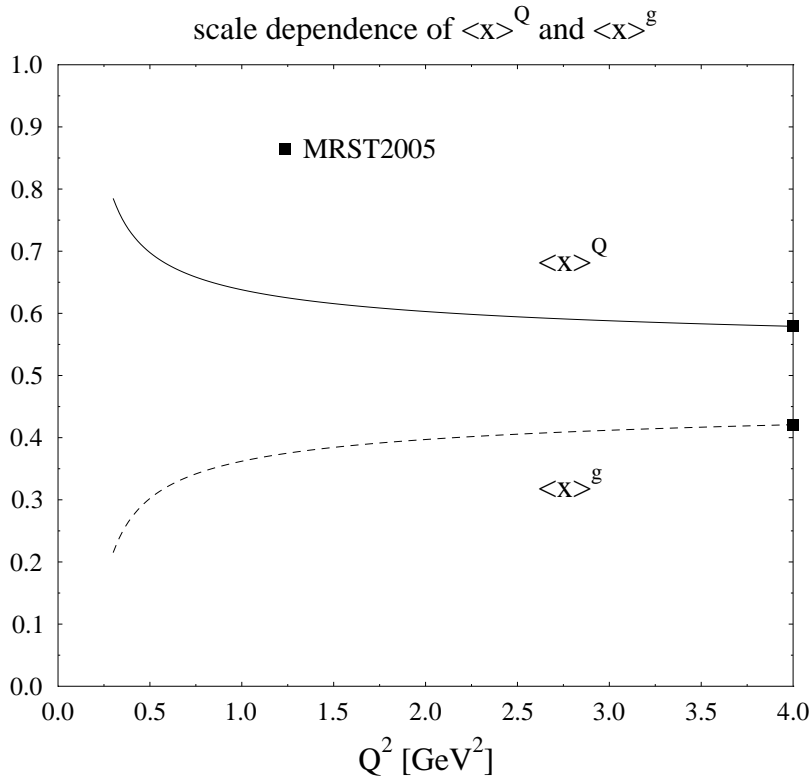}
\end{center}
\vspace*{-0.5cm}
\renewcommand{\baselinestretch}{1.20}
\caption{The scale dependencies of the quark and gluon momentum
fractions, which reproduce the MRST fits at $Q^2 = 4 \,\mbox{GeV}^2$.}
\label{Fig2:xQg}
\end{figure}

After explaining our general strategy, let us now start our
semi-phenomenological analysis of the nucleon spin contents.
We start with the empirical information obtained from the
MRST2004 as well as the CTEQ5 fits \cite{MRST04},\cite{CTEQ5}.
As already emphasized, these
two popular PDF fits give almost the same quark and gluon momentum
fractions below the energy scale $Q^2 \simeq 10 \,\mbox{GeV}^2$,
Although these PFDs are given basically above $Q^2 \simeq 1 \,\mbox{GeV}^2$,
we try to see what happens if we evolve down these fits to lower
energy scale as $Q^2 \simeq 0.30 \,\mbox{GeV}^2 \simeq 
(600 \,\mbox{MeV})^2$, which is understood to be the energy scale of
the CQSM. Using the known NLO evolution equations for
$\langle x \rangle^Q$ and $\langle x \rangle^g$, together
with the MRST2004 predictions \cite{MRST04},
\begin{equation}
  \langle x \rangle^Q \simeq 0.578, \ \ \ 
  \langle x \rangle^g \simeq 0.422 \ \ \ \mbox{at} \ \ \ 
  Q^2 = 4 \,\mbox{GeV}^2 ,
\end{equation}
we have estimated the scale dependencies of
$\langle x \rangle^Q$ and $\langle x \rangle^g$
in the range $0.30 \,\mbox{GeV}^2 \le Q^2 \le 4 \,\mbox{GeV}^2$.
The result is shown in Fig.\ref{Fig2:xQg}.
One sees that the scale dependencies of
the quark and gluon momentum fractions are fairly strong below
$Q^2 \simeq 1 \,\mbox{GeV}^2$. At the low energy scale around
$Q^2 \simeq 0.30 \,\mbox{GeV}^2$, one finds that the momentum fraction
carried by the quarks is nearly $80 \%$, while that of the gluons
is about $20 \%$. 

As a matter of course, the standard view is that the applicability range
of the perturbative QCD is at least above $1 \,\mbox{GeV}$, so that one
might be a little suspicious of the physical significance of such
``dis-evolution'' to low energy scales.
Still, we believe it meaningful by the following reason.
Basically, we are following the spirit of PDF fits
by Gl\"{u}ck, Reya and Vogt \cite{GRV95},\cite{GRV96}.
As is well known, these authors
start the QCD evolution at the exceptionally low energy scales, i.e.
$Q_{ini}^2 \simeq 0.23 \,\mbox{GeV}^2$ in the leading-order (LO) case,
and $Q_{ini}^2 \simeq 0.34 \,\mbox{GeV}^2$ in the NLO case. They
thus found that, even at such low energy scales, they absolutely need
nonperturbatively (or dynamically) generated sea-quarks, which may be
interpreted as the effects of meson clouds. We believe such
analyses (somewhat nonstandard from the viewpoint of more
conservative use of the perturbative QCD) play an
important role to connect the physics of nonperturbative QCD
in the low energy domain and the perturbative QCD in the
high-energy DIS domain.
In fact, we have carried out several theoretical analyses based on
the GRV spirit. That is, we use the predictions of the CQSM for
various PDFs as initial-scale distributions given at the low energy
scale around $600 \,\mbox{MeV}$. After evolving them with use of the
NLO evolution equation, we compare the resultant predictions with the
corresponding DIS observables with a remarkable success without
any other adjustable parameters \cite{WK99},\cite{Waka03},
\cite{WGR96}\nocite{GRW98}\nocite{DPPPW96}-\cite{DPPPW97}.
Then, we shall continue our analysis by accepting the viewpoint that
the energy scale between $600 \mbox{MeV}$ and $1 \,\mbox{GeV}$
is an important region, which connects the low energy nonperturbative
physics and the high energy perturbative physics of QCD.

\begin{figure}[htb] \centering
\begin{center}
 \includegraphics[width=14.0cm]{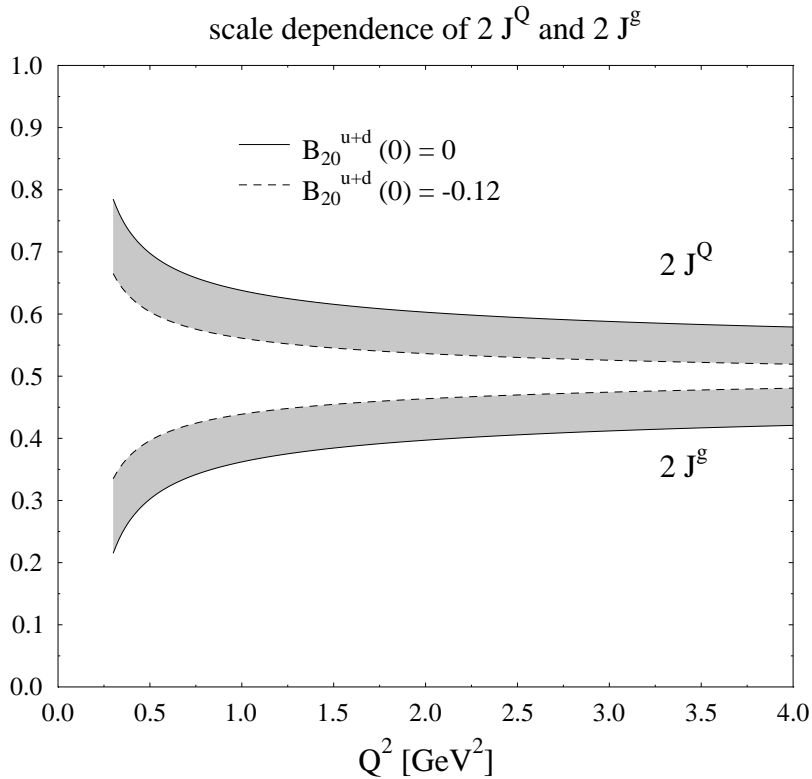}
\end{center}
\vspace*{-0.5cm}
\renewcommand{\baselinestretch}{1.20}
\caption{The scale dependencies of the quark and gluon angular momentum
fractions. The solid and dashed curves respectively correspond to
the choices $B_{20}^{u+d}(0) = 0$ and $B_{20}^{u+d}(0) = - \,0.12$.}
\label{Fig3:Jqg}
\end{figure}%

Now, we show in Fig.\ref{Fig3:Jqg} our estimate of the scale dependence
of the total angular momentum fractions carried by the quarks and
the gluons. They are obtained in the following way.
As argued in \cite{WakaNaka06}, if the net quark contribution
to the nucleon AGM vanishes, i.e. $B_{20}^{u+d}(0) = 0$, we
have extremely simple proportionality relations as
\begin{equation}
 J^Q \ = \ \frac{1}{2} \,\langle x \rangle^Q, \ \ \ \ 
 J^g \ = \ \frac{1}{2} \,\langle x \rangle^g ,
\end{equation}
which was advocated by Teryaev based on the equivalence principle some
years ago \cite{Teryaev99}\nocite{Teryaev98}-\cite{Teryaev03}.
Very interestingly, these proportionality relations hold
independently of the energy scale, since $(J^Q, J^g)$ and
$(\langle x \rangle^Q, \langle x \rangle^g)$ obey the same evolution
equations. Thus, the solid curves in Fig.\ref{Fig3:Jqg} is nothing
different from the curves for $\langle x \rangle^Q$ and
$\langle x \rangle^g$ in Fig.\ref{Fig2:xQg}.
On the other hand, the dashed curves correspond to another
extreme, which is obtained by using the value
$B_{20}^{u+d}(0) = - \,0.12$ at $Q^2 = 0.30 \,\mbox{GeV}^2$.
With this negative value of $B_{20}^{u+d}(0)$, $2 \,J^Q$ becomes
a little smaller and $2 \,J^g$ becomes a little larger as compared
with the case $B_{20}^{u+d}(0) = 0$. Still, one notices that,
at the scale $Q^2 \simeq 0.30 \,\mbox{GeV}^2$, the quarks carry about
$65 \%$ of the total angular momentum fraction. At the moment, there is
a sizable ambiguity in the magnitude of $B_{20}^{u+d}(0)$, but we
believe that the truth lies between the two extreme cases illustrated
in Fig.\ref{Fig3:Jqg}. (See the discussion at the end of the
previous section.)

Now, the net orbital angular momentum fractions carried by the
quarks can be obtained by subtracting $\Delta \Sigma$ from $2 \,J^Q$.
Since the prediction of the CQSM at $Q^2 \simeq 5 \,\mbox{GeV}^2$
is remarkably close to the central value $\Delta \Sigma = 0.33$ of
the recent HERMES analysis, we use this HERMES value here.
To make the discussion simple, we shall neglect here the scale
dependence of $\Delta \Sigma$. (In the $\overline{MS}$ scheme at the
NLO, $\Delta \Sigma$ is known to have a weak scale dependence
due to the coupling with $\Delta g$. This scale dependence is very
weak, however.) 

\begin{figure}[htb] \centering
\begin{center}
 \includegraphics[width=14.0cm]{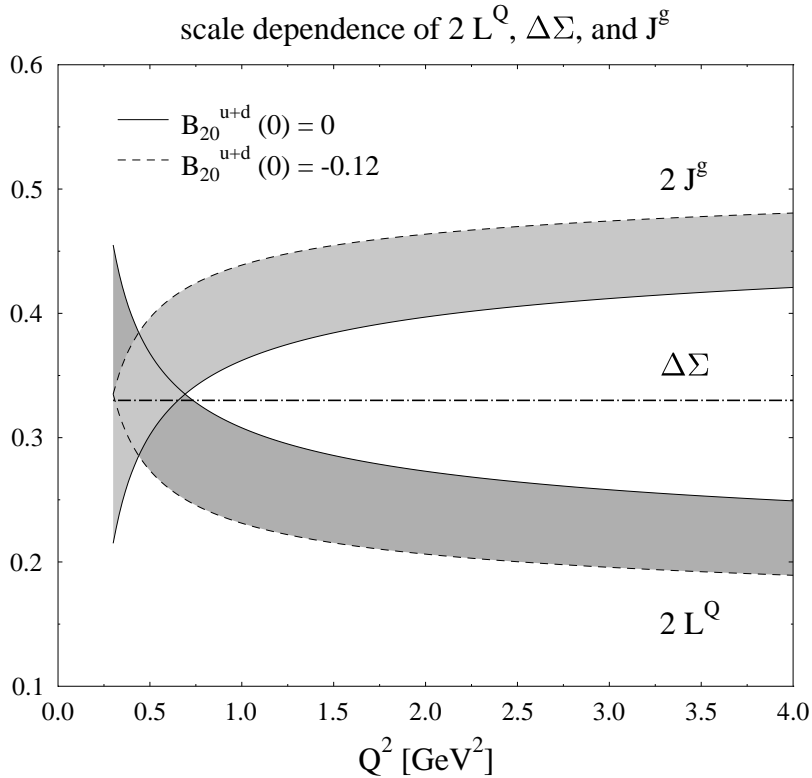}
\end{center}
\vspace*{-0.5cm}
\renewcommand{\baselinestretch}{1.20}
\caption{The scale dependencies of the total angular momentum and
orbital angular momentum of quarks, together with the net longitudinal
polarization of quarks.
The solid and dashed curves respectively correspond to
the choices $B_{20}^{u+d}(0) = 0$ and $B_{20}^{u+d}(0) = - \,0.12$}
\label{Fig4:LsqJg}
\end{figure}%

Shown in Fig.\ref{Fig4:LsqJg} are
$2 \,L^Q$, $\Delta \Sigma$, and $2 \,J^g$ as functions of
$Q^2$. Here, the solid and dashed curves correspond to the cases
$B_{20}^{u+d}(0) = 0$ and $B_{20}^{u+d}(0) = - \,0.12$,
respectively. First, let us look into the case $B_{20}^{u+d}(0) = 0$.
In this case, there is a cross over around $Q^2 \simeq 0.7 \,
\mbox{GeV}^2 \simeq (840 \,\mbox{MeV})^2$, where the magnitudes of
$2 \,L^Q$, $\Delta \Sigma$, and $2 \,J^g$ are all approximately
equal,
\begin{equation}
 2 \,L^Q \ \simeq \ \Delta \Sigma \ \simeq \ 2 \,J^g \ \simeq \ 1/3,
\end{equation}
One sees that $L^Q$ is a rapidly decreasing function of
$Q^2$, so that, as $Q^2$ increases beyond this cross over energy
scale, $L^Q$ becomes less and less important as compared with
$\Delta \Sigma$ and $J^Q$. However, the fact that $L^Q$ is a
rapidly decreasing function below $1 \,\mbox{GeV}$
conversely means that it must be very large at the low energy
scale around $600 \,\mbox{MeV}$, which we emphasize is qualitatively
consistent with the picture of the CQSM \cite{WY91},\cite{WW00}.

Next, we turn to the case $B_{20}^{u+d}(0) = - \,0.12$.
In this case, the crossover, where 
$2 \,L^Q \ \simeq \ \Delta \Sigma \ \simeq \ 2 \,J^g \ \simeq \ 1/3$,
occurs around the energy scale $Q^2 \simeq 0.30 \,\mbox{GeV}^2$.
Although the role of quark OAM is
less important as compared with the case corresponding to
$B_{20}^{u+d}(0) = 0$, it still carries about 1/3 of the total
nucleon spin at this low energy scale.
We emphasize that this is an inevitable conclusion
of believing the QCD evolution equation, since it tells us that,
at this low energy scale, the gluon (spin plus OAM) carries
at most 1/3 of the total nucleon spin, so that what remains to
carry the rest 1/3 of the nucleon spin must be the quark OAM. 
On the other hand, when going to higher energy scale, say
at $Q^2 \simeq 4 \,\mbox{GeV}^2$, corresponding to the
renormalization scale of the lattice QCD calculations,
one sees that the amount of the quark OAM becomes much
smaller. Still, it is seen to carry nearly $20 \%$ of the
total nucleon spin even at $Q^2 \simeq 4 \,\mbox{GeV}^2$.
One might suspect that this would contradict the conclusion of
the lattice QCD analyses.
Probably, the main cause of discrepancy can be traced back
to a little overestimation of the net quark polarization
$\Delta \Sigma$ in the lattice QCD.
In fact, the results of the QCDSF-UKQCD group for
$\Delta \Sigma$ is $0.402 \pm 0.024$ \cite{QCDSF-UKQCD07},
which overestimate a little the central value 0.33 of the HERMES
analysis \cite{HERMESD06}, which we have used
in our semi-empirical analysis here.
(One of the reason of a little overestimation of $\Delta \Sigma$
in the lattice QCD simulations may be attributed to the so-called
quenched approximation, i.e. the neglect of the disconnected diagrams.)
In our opinion, the quark OAM fraction of the order of $20 \%$
is reasonable enough from the following simple consideration.
To convince it, we recall that the asymptotic value (the value
in the $Q^2 \rightarrow \infty$ limit) of the total angular
momentum fractions of the quarks and the gluons are extracted
from the relations (\ref{LOJQ}) and (\ref{LOJg}), which follows
from the fact that
$(J^Q, J^g)$ and $(\langle x \rangle^Q, \langle x \rangle^g)$
obey the same evolution equation. With the realistic case of
6 flavors, we have
\begin{equation}
 2 \,J^Q (\infty) \ \simeq \ 0.529, \ \ \ \ \ \ 
 2 \,J^g (\infty) \ \simeq \ 0.471.
\end{equation}
Subtracting $\Delta \Sigma \simeq 0.33$, which is thought to be nearly
scale-independent, we thus obtain
\begin{equation}
 2 \,L^Q (\infty) \ \simeq \ 0.199.
\end{equation}
Since $L^Q$ is a decreasing function of $Q^2$, the magnitude
of $2 \,L^Q$ at the scale $Q^2 \simeq 4 \,\mbox{GeV}^2$ must
be larger or at least approximately equal to this asymptotic
value, which justifies our reasoning above.  

So far, we have concentrated on the analysis of the net quark and gluon 
contribution to the nucleon spin and the net quark contribution to the 
orbital angular momentum. Now, we try to make a flavor 
decomposition of the quark contribution to the nucleon spin and orbital 
angular momentum, which requires the knowledge of the quantity 
$B_{20}^{u - d} (0),$ i.e. the isovector nucleon AGM.
Since we want to investigate the scale dependencies of the momentum 
fractions and the total angular momenta of the quarks and gluons up to
$Q^2 = 4 \,\mbox{GeV}^2$, we use again the NLO evolution equation with
$3$ active flavors, although we assume that the strange quarks carry
negligible momentum fraction and AGM at the initial low energy scale,
for simplicity.
As initial conditions of evolution, we need the following quantities at 
$Q^2 = 0.30 \,\mbox{GeV}^2$ : 
\begin{eqnarray}
 \langle x \rangle^{(0)} &=& \langle x \rangle^{u + d + s} 
 \ \equiv \ \langle x \rangle^Q, \ \ \ \ \ \ 
 \langle x \rangle^g, \\
 \langle x \rangle^{(3)} &=& \langle x \rangle^{u - d}, \hspace{10mm}
 \langle x \rangle^{(8)} \ = \ \langle x \rangle^{u + d - 2 s} .
\end{eqnarray}
The singlet moments $\langle x \rangle^Q$ and $\langle x \rangle^g$
evolve according to the evolution equation (\ref{NLOevolsing}).
Here, we use the initial condition
\begin{eqnarray}
 \langle x \rangle ^{u + d} &=& 0.785, \ \ \ \ 
 \langle x \rangle ^{u - d} \ = \ 0.250, \ \ \ \ 
 \langle x \rangle ^{g} \ = \ 0.215,
\end{eqnarray}
with $\langle x \rangle^s= 0$, since it gives at
$Q^2 = 4 \,\mbox{GeV}^2$
\begin{eqnarray}
 \langle x \rangle ^{u + d + s} &=& 0.579, \ \ \ \ \ 
 \langle x \rangle ^g = 0.421, \\
 \langle x \rangle ^{u + d} &=& 0.552, \ \ \ \ \ 
 \langle x \rangle ^{u - d} = 0.158 ,
\end{eqnarray}
which approximately reproduces the empirical MRST2004 fit
at the same scale \cite{MRST04}.
We can make a similar analysis also for the total angular momentum of
the quarks and gluons, because they obey the same evolution equations
as the corresponding momentum fractions.
To proceed, we need initial conditions for the following quantities
\begin{eqnarray*}
 J^{u + d} &=& \frac{1}{2} \,[ \langle x \rangle^{u + d} 
 \ + \ B_{20}^{u + d} (0) ], \\ 
 J^{u - d} &=& \frac{1}{2} \,[ \langle x \rangle^{u - d} 
 \ + \ B_{20}^{u - d} (0) ], \\ 
 J^g &=& \frac{1}{2} \,[ \langle x \rangle^g 
 \ + \ B_{20}^g (0) ],
\end{eqnarray*}
with the general constraint $B_{20}^{u + d} (0) + B_{20}^g (0) = 0$.
Here, we have assumed $J_s = 0$ at the initial scale. 
For $B_{20}^{u + d} (0)$, we consider the two cases again, i.e
$B_{20}^{u + d} (0) \, = \, 0$ and
$B_{20}^{u + d} (0) \ = \ - \,0.12$.
For $B_{20}^{u - d} (0)$, we use the prediction of the CQSM given by
\begin{equation}
 B_{20}^{u - d} (0) \ = \ 0.458 .
\end{equation}
At first sight, the magnitude of the isovector AGM above seems to be
fairly larger than the corresponding predictions of the lattice QCD
given at $Q^2 = 4 \,\mbox{GeV}^2$.
As already mentioned in the previous section, however,
after taking account of the scale dependence, we find that the above
CQSM prediction for $B_{20}^{u - d} (0)$ is remarkably close to that
of the lattice QCD. 

\begin{figure}[htb] \centering
\begin{center}
 \includegraphics[width=14.0cm]{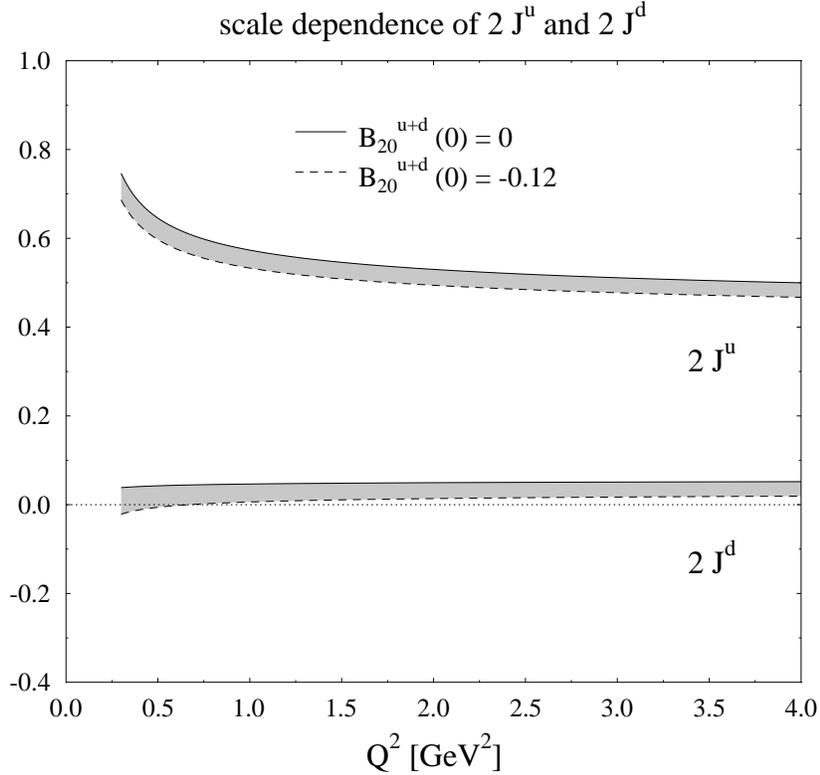}
\end{center}
\vspace*{-0.5cm}
\renewcommand{\baselinestretch}{1.20}
\caption{The scale dependencies of $J^u$ and $J^d$.
The solid and dashed curves respectively correspond to
the choices $B_{20}^{u+d}(0) = 0$ and $B_{20}^{u+d}(0) = - \,0.12$}
\label{Fig5:JuJd}
\end{figure}

Now that all the necessary conditions are given at the initial scale 
$Q^2 = 0.30 \,\mbox{GeV}^2$, let us first try to estimate the total
angular momentum fractions at $Q^2 = 4 \,\mbox{GeV}^2$, which
corresponds to the renormalization scale of lattice QCD simulations.
First, we show the results corresponding to the choice 
$B_{20}^{u + d} (0) = 0$. We have, at $Q^2 = 4 \,\mbox{GeV}^2$, 
\begin{eqnarray}
 \langle x \rangle ^Q &=& 0.579, \ \ \ \langle x \rangle ^g = 0.421, \\
 \langle x \rangle ^{u + d} &=& 0.552, \ \ \ 
 \langle x \rangle^{u - d} \ = \ 0.158 , \ \ \ 
 \langle x \rangle ^s \ = \ 0.028,
\end{eqnarray}
and
\begin{eqnarray}
 2 \,J^Q &=& 0.579, \ \ \ 2 \,J^g = 0.421, \\
 2 \,J^{u + d} &=& 0.552, \ \ \ 2 J^{u - d} \ = \ 0.448,
 \ \ \ 2 \,J^s = 0.028,
\end{eqnarray}
which gives
\begin{equation}
 2 \,J^u \ \simeq \ 0.500, \ \ \ 2 \,J^d \ \simeq \ 0.052 .
\end{equation}
On the other hand, with the choice $B_{20}^{u + d} (0) = -0.12$ 
at the initial energy scale, we get
\begin{eqnarray}
 2 \,J^Q &=& 0.519, \ \ \ 2 \,J^g \ = \ 0.481 \\
 2 \,J^{u + d} &=& 0.486, \ \ \ 2 \,J^{u - d} \ = \ 0.448,
 \ \ \ 2 \,J^s \ = \ 0.033 ,
\end{eqnarray}
which gives
\begin{equation}
 2 \,J^u \ \simeq \ 0.467, \ \ \ \ \ 2 \,J^d \ \simeq \ 0.019
\end{equation}
Depending on the two choices for $B_{20}^{u + d} (0)$. i.e. 
$B_{20}^{u + d} (0) \simeq -0.12$ or $B_{20}^{u + d} (0) = 0$, 
we thus obtain an estimate,
\begin{equation}
 2 \,J^u \ \simeq \ 0.46 - 0.50, \ \ \ \ \ 
 2 \,J^d \ \simeq \ 0.02 - 0.05 ,
\end{equation}
which supports the conclusion of the lattice QCD studies 
that the total angular momentum carried by the $d$-quarks is nearly zero
at least qualitatively. For reference, we show in Fig.\ref{Fig5:JuJd}
the predicted scale dependence of $J^u$ and $J^d$.

Now, the information on the quark OAM can be obtained from $J^u, J^d$ 
and $J^s$ by subtracting the corresponding intrinsic spin contributions.
Here, we use the empirical information provided by the recent 
HERMES analysis \cite{HERMESD06}, which gives at $Q^2 = 5 \,\mbox{GeV}^2$,
\begin{eqnarray}
 g_A^{(0)} &\equiv& \Delta \Sigma^{u + d + s} 
 = 0.330 \pm 0.011 \mbox{(theor.)} \pm 0.025 \mbox{(exp.)}
 \pm 0.028 \mbox{(evol.)}, \\
 g_A^{(3)} &\equiv& \Delta \Sigma^{u - d} = 1.269 \pm 0.003, \\
 g_A^{(8)} &\equiv& \Delta \Sigma^{u + d - 2 s} = 0.586 \pm 0.031,
\end{eqnarray}
Neglecting the error-bars, for simplicity, this gives
\begin{eqnarray}
 \Delta \Sigma^u &=& 0.842, \ \ \ 
 \Delta \Sigma^d \ = \ - \,0.427, \ \ \ 
 \Delta \Sigma^s \ = \ - \,0.085 .
\end{eqnarray}
As is well known, due to the conservation of the flavor nonsinglet
axial-current, $g_A^{(3)}$ and $g_A^{(8)}$ are exactly scale independent.
Then, if we neglect very weak scale-dependence of $g_A^{(0)}$, 
all of $\Delta \Sigma^u, \Delta \Sigma^d$ and $\Delta \Sigma^s$ are
thought to be scale independent.
Let us first estimate the quark OAM at $Q^2 = 4 \,\mbox{GeV}^2$.
Depending on the two choices for $B_{20}^{u + d} (0)$, i.e. 
$B_{20}^{u + d}(0) = - \,0.12$ and $B_{20}^{u + d} (0) = 0$, we obtain
\begin{eqnarray}
 2 L^u &=& - \,(0.333 - 0.300), \ \ \ 
 2 L^d \ = \ 0.489 - 0.522, \ \ \ 
 2 L^s \ = \ 0.033 - 0.028 .
\end{eqnarray}
A prominent feature here is that the magnitudes of $L^u$ and $L^d$ are 
sizably large with the opposite sign such that $L^u < 0$ and $L^d > 0$,
which leads to the inequality
\begin{equation}
 | L^{u + d} | \ \ll \ | L^{u - d} |,
\end{equation}
i.e.~the {\it isovector dominance} of the quark OAM.
(As already discussed, the 
cancellation between $L^u$ and $L^d$ is not so perfect in our 
semi-phenomenological analysis as compared with the lattice QCD 
predictions.) To understand the physical meaning of the above unique
feature, we find it instructive to look into the scale dependence
of $L^{u-d}$ as well as of $L^u$ and $L^d$.

\begin{figure}[htb] \centering
\begin{center}
 \includegraphics[width=14.0cm]{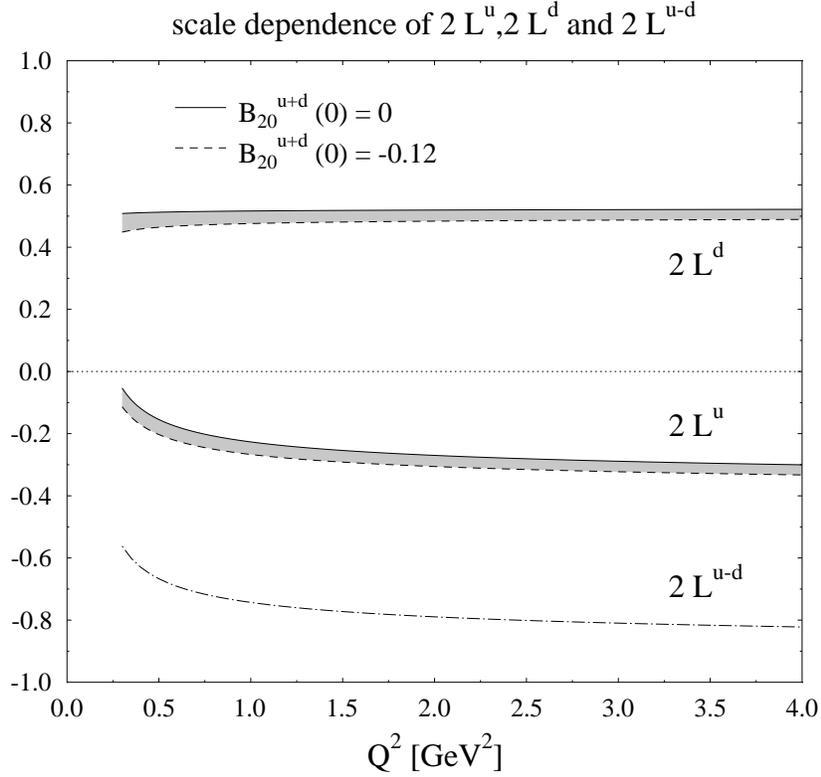}
\end{center}
\vspace*{-0.5cm}
\renewcommand{\baselinestretch}{1.20}
\caption{The scale dependencies of $L^u$ and $L^d$ as well as $L^{u-d}$.
The solid and dashed curves respectively correspond to
the choices $B_{20}^{u+d}(0) = 0$ and $B_{20}^{u+d}(0) = - \,0.12$}
\label{Fig6:LuLd}
\end{figure}%

Shown in Fig.\ref{Fig6:LuLd} are scale dependencies of
$2 L^u, 2 L^d$, and $ 2 L^{u - d}$.
(Note that the difference $L^{u - d}$ of $L^u$ and $L^d$ does not depend 
on the choice of $B_{20}^{u + d} (0).$)
One clearly sees that $L^{u - d}$ is a decreasing function of 
$Q^2$. Since $L^{u - d}$ is negative, this means that $|L^{u - d}|$ is 
an increasing function of $Q^2$.
Actually, this somewhat peculiar behavior of $L^{u-d}$ can
naturally be understood from the definitional equation of quark OAM : 
\begin{equation}
 2 \,L^{u - d} (Q^2) \ \equiv \ 2 \,J^{u - d} (Q^2) \ - \ 
 \Delta \Sigma^{u - d} .
\end{equation}
Since $J^{u - d}(Q^2)$ is a decreasing function of $Q^2$, while
$\Delta \Sigma^{u - d}$ is $Q^2$-independent, $L^{u - d} (Q^2)$ is a 
decreasing function of $Q^2$. In particular, since 
$2 J^{u - d} (\infty) = 0$, as verified from the non-singlet (NS)
evolution equation (\ref{NLOevolNS}), one finds that the isovector
quark OAM in the asymptotic limit $Q^2 \rightarrow \infty$ is solely
determined by the isovector axial-charge of the nucleon 
$g_A^{(I = 1)} \equiv \Delta \Sigma^{u - d}$ as
\begin{equation}
 2 \,L^{u - d} (\infty) \ = \ - \,\,g_A^{(I = 1)} \ = \ - \,1.269 .
\end{equation}
This is really an astonishing observation, since it means that the quark
OAM in the asymptotic limit, at least its isovector combination,
is determined solely by the longitudinal quark 
polarization ! Note that, since there is no room for doubt in 
using the relation $L^q = J^q - \frac{1}{2} \Delta \Sigma^q$ to extract
quark OAM, this mysterious conclusion is an inevitable consequence of
the following two theoretical postulates :

\begin{itemize} 
\item the definition of $J^q$ through Ji's angular momentum sum rule,
$J^q = \frac{1}{2} \,[\langle x \rangle^q + B_{20}^q (0) ]$.
\item the observation that $J^q$ and $\langle x \rangle^q$ obey the
same evolution equation.
\end{itemize}

Anyhow, since the net quark OAM $2 L^{u + d} (Q^2)$ is a rapidly
decreasing function of $Q^2$, we can easily understand the
feature that $L^u$ is large
and negative, while $L^d$ is large and positive above a few GeV scale.
It is an interesting open question whether such a large OAM of the
$u$- and $d$-quarks with opposite sign can be verified through some
direct measurements like the single-spin asymmetry of semi-inclusive
reactions depending on the Sivers mechanism \cite{Sivers90},
which is believed to be sensitive to the OAM of nucleon
constituents.

Now, we attempt to give a complete solution to
our first question, i.e. the problem of determining
the full spin contents of the nucleon. Our answer on the decomposition
of the nucleon spin into the sum of $L^Q$, $\frac{1}{2} \,\Delta \Sigma$,
and $J^g$ is already given in Fig.\ref{Fig4:LsqJg} within the range
$0.30 \,\mbox{GeV}^2 \lesssim Q^2 \lesssim 4.0 \,\mbox{GeV}^2$.
Still, uncompleted is further decomposition of $J^g$ into
the sum of $\Delta g$ and $L^g$. Unfortunately, this decomposition
is not gauge invariant and it cannot be done very reliably
as compared with the analysis done so far.
Still, the following qualitative consideration would be of some
help to have a rough idea about the complete decomposition of the nucleon
spin, thereby clarifying the fairly confused situation pointed out
in Introduction. A basis of the following analysis is the observation
that the gluon polarization
in the nucleon cannot be very large at least at the low renormalization
scales \cite{COMPASSD05},\cite{COMPASSD06} and HERMES
groups \cite{HERMESD06}.
As a simplest trial, we therefore
assume that the gluon polarization $\Delta g$ is zero, at the low
energy model scale around $Q^2 = 0.3 \,\mbox{GeV}^2$.
Combining this with the CQSM prediction $\Delta \Sigma = 0.35$ for
the net quark longitudinal polarization, we solve the NLO evolution
equation for $\Delta \Sigma$ and $\Delta g$ in the standard
$\overline{\rm MS}$ factorization scheme.

\begin{figure}[htb] \centering
\begin{center}
 \includegraphics[width=14.0cm]{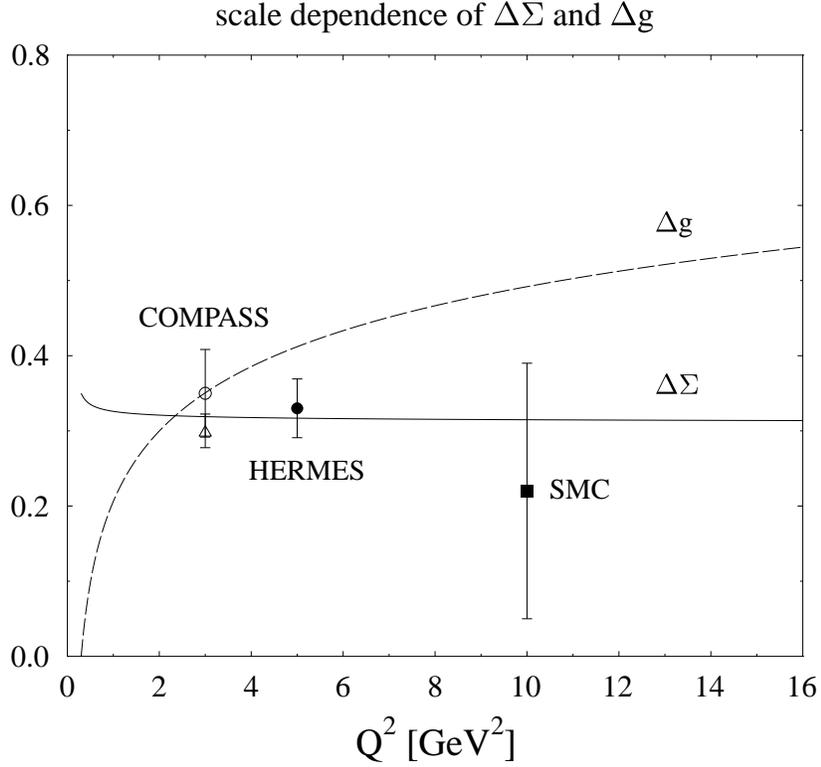}
\end{center}
\vspace*{-0.5cm}
\renewcommand{\baselinestretch}{1.20}
\caption{The scale dependencies of $\Delta \Sigma$ and $\Delta g$, obtained
as explained in the text, are compared with the recent QCD fits by the
COMPASS group (open circle and open triangle) and by the HERMES
group (filled circle). The old SMC result is also shown for reference
by the filled square.}
\label{Fig7:DelSdelg}
\end{figure}%

The resultant $\Delta \Sigma$ and $\Delta g$ as functions of $Q^2$ are
illustrated in Fig.\ref{Fig7:DelSdelg} together with the empirical
values obtained in the recent NLO analyses by the COMPASS and the HERMES
groups \cite{COMPASSD05}\nocite{COMPASSD06}-\cite{HERMESD06},
as well as the old SMC fit \cite{SMC98}.
As repeatedly emphasized, the new COMPASS and the HERMES
results for $\Delta \Sigma$ are remarkably close to the prediction of
the CQSM.
Also noteworthy here is the strong scale dependence of the longitudinal
gluon polarization. In spite that we have assumed that $\Delta g$
is zero at the starting energy scale, it grows rapidly with increasing
$Q^2$. As nicely explained in \cite{Cheng98}, the growth of the gluon
polarization with $Q^2$ can be traced back to the positive sign of
the relevant anomalous dimension $\delta \gamma^{(0)1}_{qg}$.
The positivity of this quantity dictates that the polarized quark is
preferred to radiate a gluon with helicity parallel to the quark
polarization. Since the net quark spin component in the proton is
clearly positive, it follows that $\Delta g > 0$ at least for the
gluon perturbatively radiated from the quarks. The growth rate of
$\Delta g$ is so fast especially in the relatively low $Q^2$ region
that its magnitude reaches around $(0.3 - 0.4)$ already at
$Q^2 = 3 \,\mbox{GeV}^2$, which may be compared with the estimate
given by the COMPASS group :
\begin{equation}
 \Delta g (Q^2 = 3 \,\mbox{GeV}^2)_{COMPASS} \ \simeq \ 
 (0.2 - 0.3) .
\end{equation}
It should be emphasized that the gluon polarization of this size is
nothing inconsistent with the GRSV standard scenario of the polarized
PDF fit \cite{GRSV01}. (Almost the same viewpoint was emphasized also
in a recent bag model study of the gluon polarization \cite{CJ06}.)
Let us therefore proceed further by assuming that our estimate of
$\Delta g$ shown in Fig.\ref{Fig7:DelSdelg} is not extremely far from
the reality, which enables us to carry out a decomposition of
$J^g$ into $\Delta g$ and $L^g$.

\begin{figure}[htb] \centering
\begin{center}
 \includegraphics[width=14.0cm]{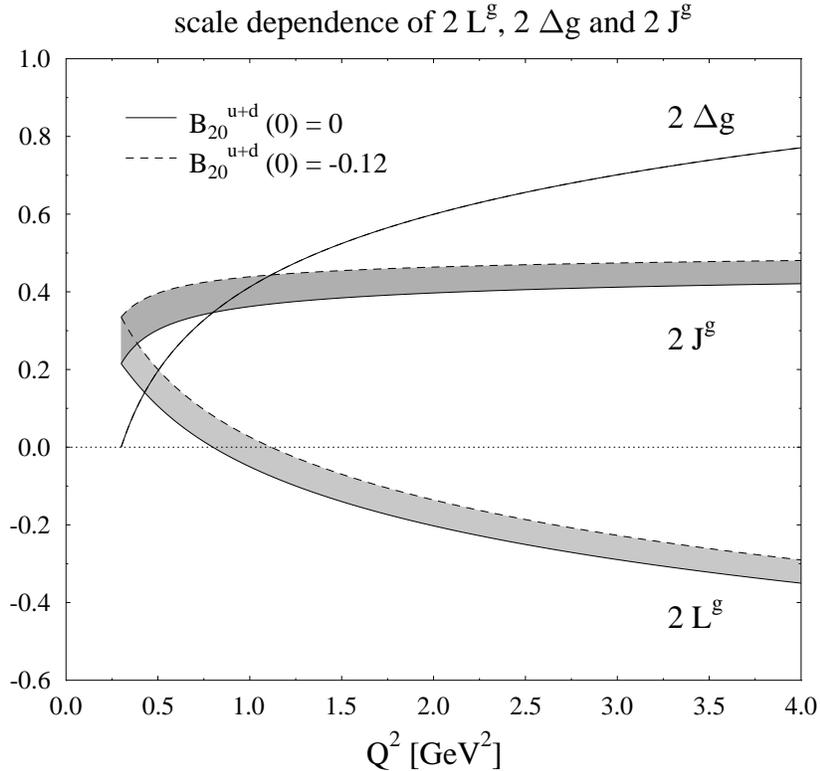}
\end{center}
\vspace*{-0.5cm}
\renewcommand{\baselinestretch}{1.20}
\caption{The spin and OAM decomposition of gluon total angular
momentum as a function of $Q^2$.}
\label{Fig8:Lg}
\end{figure}%

Fig.\ref{Fig8:Lg} shows the gluon OAM $L^g$ obtained in the above way,
together with $2 \,J^g$ and $2 \,\Delta g$.
One sees that the gluon OAM $L^g$ is a rapidly decreasing function
of $Q^2$. This feature naturally follows since
$L^g = J^g - \Delta g$ and the increasing rate of $\Delta g$ is
much faster than that of $J^g$.
Very interestingly, the magnitude of $L^g$ in the vicinity
$Q^2 \simeq 1 \,\mbox{GeV}^2$ turns out to be fairly close to zero.
We are not sure whether this can be interpreted as giving
a support to Brodsky and Gardner's interpretation of the recent COMPASS
observation of small single-spin asymmetry on the isoscalar
deuteron target.

Anyhow, keeping in mind that the spin decomposition of the nucleon is
highly scale dependent, our estimate at the scale
$Q^2 \simeq 4 \,\mbox{GeV}^2$ can be summarized as follows.
The net angular momentum
fractions carries by the quarks and the gluons are 
$2 \,J^Q \simeq 0.52 - 0.58$ and $2 \,J^g \simeq 0.42 - 0.48$.
The total angular momentum carried by the quarks can further be
decomposed into the spin and OAM parts as $\Delta \Sigma \simeq 0.33$
and $2 \,L^Q \simeq 0.19 - 0.25$. The decomposition of $J^g$ into
the sum of $\Delta g$ and $\L^g$ is still very ambiguous.
But, the standard scenario for the evolution of $\Delta g$ indicates
that the gluon total angular momentum of the order of
$2 \,J^g \simeq 0.42 - 0.48$ is a consequence of the cancellation of
relatively large and positive $\Delta g$ and negative gluon OAM with
a little smaller magnitude.

Finally, we make a short comment on the recent extraction of the quark
total angular momentum through the model-dependent GPD analyses
of the semi-inclusive reactions.
The first experimental result for the quark angular momentum was
obtained by the HERMES Collaboration by studying the hard exclusive
$\pi^0$ production on the transversely polarized hydrogen
target \cite{HERMESJ05}.
Their results, corresponding to the average energy scale
$Q^2 \simeq 2.5 \,\mbox{GeV}^2$, is given
by \cite{HERMESJ07},\cite{HERMESJ06}
\begin{equation}
 J^u + J^d / 2.9 \ = \ 0.42 \ \pm \ 0.21 \ \pm \ 0.06 .
\end{equation}
On the other hand, another combination of $J^u$ and $J^d$ was
extracted by the JLab Hall A Collaboration through the analysis
of the DVCS and the Bethe-Heitler processes on the neutron and
on the deuteron target \cite{JLabHA07}.
Their result, corresponding to the average energy scale $Q^2 \simeq 
1.9 \,\mbox{GeV}^2$, is given by
\begin{equation}
 J^d + J^u / 5.0 \ = \ 0.18 \ \pm 0.14 .
\end{equation}
For reference, we show below the corresponding predictions of our
semi-phenomenological analysis.
Depending on the two choices $B_{20}^{u+d}(0) = - \,0.12$ and
$B_{20}^{u+d}(0) = 0$, we obtain
\begin{equation}
 J^u + J^d / 2.9 \ = \ (0.245 - 0.268), \hspace{10mm}
 Q^2 = 2.5 \,\mbox{GeV}^2 ,
\end{equation}
and
\begin{equation}
 J^d + J^u / 5.0 \ = \ (0.056 - 0.078), \hspace{10mm}
 Q^2 = 1.9 \,\mbox{GeV}^2 .
\end{equation}
Clearly, our estimates lie in the allowed ranges of both the HERMES and
JLab determinations of $J^u$ and $J^d$.
However, it is also clear that the error-bars of the two
determinations are still too large to be able to say something definite.

\section{Concluding remarks}

After completing our semi-empirical analysis of the nucleon spin
contents, we now try to answer several questions raised in
Introduction.
Accepting the observation that the intrinsic quark spin carries
about $1/3$ of the total nucleon spin, what carry the rest of it ?
As we have shown, the answer depends on the scale of observation
in an essential manner. At the relatively high energy scale around
$Q^2 \simeq 4 \,\mbox{GeV}^2$, corresponding to the renormalization
scale of the recent lattice QCD simulations, the quarks and gluons
respectively carry about $(52 - 58) \,\%$ and $(0.42 - 0.48) \,\%$
of the total nucleon spin.
The total angular momentum fraction $2 \,J^Q$
carried by the quarks can further be decomposed into the spin and
OAM parts as $\Delta \Sigma \simeq 0.33$ and $2 \,L^Q \simeq 
0.19 - 0.25$. Our estimate for the quark OAM
appears to contradict the conclusion of the lattice QCD
studies that the OAM carried the quarks is nearly
zero. The cause of discrepancy can mainly be traced back
to a little overestimation of the net longitudinal quark polarization
$\Delta \Sigma$ in the lattice QCD simulation.
In fact, once we accept to use the central value $\Delta \Sigma
= 0.33$ given by the recent HERMES fit, the quark
OAM of the order of $20 \%$ at $Q^2 \simeq 4 \,\mbox{GeV}^2$
is nothing unreasonable as can be convinced from the following two
observations made based on the evolution equations of relevant quantities.
The one is the fact that the asymptotic ($Q^2 \rightarrow \infty)$
value of the net quark OAM fraction is given by
$2 \,L^Q (\infty) = 2 \,J^Q (\infty) - \Delta \Sigma \simeq 
0.529 - 0.33 \simeq 0.199$ for $n_f =6$. The other is the fact
that $L^Q$ is a decreasing function of $Q^2$.
The decomposition of $J^g$ into the sum of the spin and the OAM
parts is still very ambiguous. Nonetheless, the standard scenario
for the evolution of $\Delta g$ strongly indicates that the total
gluon angular momentum of the order $2 \,J^g = 0.42 - 0.48$ at
$Q^2 = 4 \,\mbox{GeV}^2$ is likely to be a consequence of the
cancellation of relatively large and positive $\Delta g$ and
negative gluon OAM with a little smaller magnitude.

At the low energy scales of nonperturbative QCD around
$Q^2 \simeq (0.30 - 0.70) \,\mbox{GeV}^2$, we get a very
different picture on the nucleon spin contents. In these energy
scales, the quark OAM, the intrinsic quark spin, and the gluon total
angular momentum would give roughly the same magnitude of
contributions to the nucleon spin, i.e. $2 \,L^Q \simeq \Delta \Sigma
\simeq 2 \,J^g \simeq 1/3$. 

Also very interesting is the flavor decomposition of the total angular
momentum and the OAM carried by the quarks. On the basis of Ji's
observation that $J^q$ and $\langle x \rangle^q$ obey the same evolution
equation, we have shown that the asymptotic limit of the isovector
quark OAM is {\it solely} determined by the isovector
axial-charge of the nucleon or the isovector part of the longitudinal
quark polarization as $2 \,L^{u-d} (\infty) = - \,g_A^{(I=1)}
= - \Delta \Sigma^{u-d} = - \,1.269$, which leads to novel
{\it isovector dominance} of the quark OAM at the high energy scale.
It is an interesting open question whether this unique feature of
the quark OAM at high $Q^2$ can be probed through some direct
observations in high energy DIS processes.

\vspace{3mm}
\begin{acknowledgments}
This work is supported in part by a Grant-in-Aid for Scientific
Research from Ministry of Education, Culture, Sports, Science
and Technology, Japan (No.~C-16540253)
\end{acknowledgments}


\end{document}